\documentclass[11pt, 3p]{elsarticle}

\usepackage{mathpazo}
\usepackage[T1]{fontenc}
\usepackage[utf8]{inputenc}
\usepackage{multirow}
\usepackage[colorlinks=true]{hyperref}
\usepackage[english]{babel}
\usepackage{adjustbox}
\usepackage{xcolor}
\usepackage{booktabs}
\usepackage{amsmath}
\usepackage{textcomp}
\usepackage{lineno}
\usepackage{graphicx}
\usepackage{tikz}
\usepackage{siunitx}
\usepackage{setspace}
\usepackage{natbib}
\usepackage[caption=false]{subfig}
\usepackage{textcomp}
\usepackage{lipsum}
\usepackage{caption}
\usepackage[font=small,labelfont=bf, skip=0pt]{caption}
\biboptions{sort&compress}

\usepackage{soulutf8}
\biboptions{sort&compress}

\graphicspath{{./}} 

\begin{document}
%\nocite{TitlesOn}
	
\begin{frontmatter}

\journal{Journal of Alloys and Compounds}

\title{Evidence for multiband superconductivity and charge density waves in Ni-doped ZrTe$_2$}
 
\author[1]{Lucas E. Correa}
\author[1]{Pedro P. Ferreira\corref{cor}}
\cortext[cor]{Corresponding author, website: \url{https://computeel.org}}
\ead{pedroferreira@usp.br}
\author[1]{Leandro R. de Faria}
\author[2]{Thiago T. Dorini}
\author[3]{Mário S. da Luz}
\author[4]{Zachary Fisk}
\author[5]{Milton S. Torikachvili}
\author[1]{Luiz T. F. Eleno}
\author[1]{Antonio J. S. Machado\corref{cor}}
\ead{ajefferson@usp.br}

\address[1]{Universidade de S\~ao Paulo, Escola de Engenharia de Lorena, Departamento de Engenharia de Materiais, Lorena, S\~ao Paulo, Brasil}
\address[2]{Universit\'e de Lorraine, CNRS, IJL, Nancy, France}
\address[3]{Instituto de Ci\^encias Tecnol\' ogicas e Exatas, Universidade Federal do Tri\^angulo Mineiro, Uberaba, Minas Gerais, Brasil}
\address[4]{University of California, Irvine, California 92697, USA}
\address[5]{Department of Physics, San Diego State University, San Diego, California 92182-1233, USA}
\begin{abstract}

%\linenumbers

	We carried out a comprehensive study of the electronic, magnetic, and thermodynamic properties of Ni-doped ZrTe$_2$. High quality Ni$_{0.04}$ZrTe$_{1.89}$ single crystals show a possible coexistence of charge density waves (CDW, T$_{CDW}\approx287$\,K) with superconductivity (T$_c\approx 4.1$\,K), which we report here for the first time. The temperature dependence of the lower (H$_{c_1}$) and upper (H$_{c_2}$) critical magnetic fields both deviate significantly from the behaviors expected in conventional single-gap s-wave superconductors. However, the behaviors of the normalized superfluid density $\rho_s(T)$ and H$_{c_2}(T)$  can be described well using a two-gap model for the Fermi surface, in a manner consistent with conventional multiband superconductivity. Electrical resistivity and specific heat measurements show clear anomalies centered near 287\,K consistent with a CDW phase transition. Additionally, electronic-structure calculations support the coexistence of electron-phonon multiband superconductivity and CDW order due to the compensated disconnected nature of the electron- and hole-pockets at the Fermi surface. Our electronic structure calculations also suggest that ZrTe$_2$ could reach a non-trivial topological type-II Dirac semimetallic state. These findings highlight that Ni-doped ZrTe2 can be uniquely important for probing the coexistence of superconducting and CDW ground states in an electronic system with non-trivial topology.
	
\end{abstract}


\end{frontmatter}

%\resetlinenumber
%\linenumbers

\section{Introduction}

Transition metal dichalcogenides (TMDs) are layered quasi-2 dimensional compounds of general composition MX$_2$, where M is a transition metal (e.g. Zr, Hf, Ti and Ta) and X is a chalcogen (S, Se and Te) \cite{manzeli2017, wang2012, chhowalla2013}. These materials have drawn a heightened level of attention recently due to a plethora of uncommon physical properties, including the coexistence of superconductivity and charge density wave (CDW) and emergent quantum states \cite{neto2001, valla2004, morosan2006, sipos2008, zhou2016, bhoi2016, han2018}. The latter are particularly relevant for applications including quantum information, spintronic devices and battery systems \cite{zibouche2014, xia2014, yang2015, wu2017, krasnok2018, david2019, lucatto2019}.

Individual layers are composed of three atomic planes with X-M-X stacking. Bulk samples and heterostructures are formed by stacked layers bound together by van der Waals interactions. Therefore, they can show different behaviors according to the stacking order and metal coordination. In general, TMDs crystallize with hexagonal or rhombohedral symmetries, space-groups (SG) $P6_3/mmc$, $P\overline{3}m1$ and  $R3m$, and the M site has octahedral or trigonal prismatic coordination with the strongly bonded X layers. Such configuration is a friendly environment for the intercalation of other atomic species and complexes in the region between adjacent chalcogen planes (van der Waals gap), permitting tweaking and fine-tuning of the host’s electronic ground state \cite{wagner2008, morosan2010, kiswandhi2013, chang2016, guzman2017, kitou2019}.

The TMD ZrTe$_2$ crystallizes in a trigonal structure (CdI$_2$-type, SG $P\overline{3}m1$, number 164). The electronic behavior is typical of semimetals, though detailed studies are lacking \cite{de1997, reshak2004, kar2020}. Recent studies using
high-resolution angle-resolved photoemission spectroscopy (ARPES) and density functional theory (DFT) revealed details of the Fermi surface that suggest the possibility of a non-trivial topological order due to spin-orbit coupling induced band inversion between Zr-$d$ and Te-$p$ states at the $\Gamma$ point \cite{kar2020, tsipas2018}. In contrast, through nuclear magnetic resonance (NMR) measurements supported by DFT calculations Tian et\,al. \cite{tian2020} proposed that ZrTe$_2$ is a quasi-2D topological nodal-line semimetal \cite{tian2020}, then raising a question about the real topological nature of ZrTe$_2$.

An example of the striking effect of intercalation in the van der Waals gap is the addition of small amounts of copper in ZrTe$_2$, yielding superconductivity with T$_c\approx9.7$\,K \cite{machado2017}. The superconducting behavior deviates from BCS and it is consistent with multiband superconductivity. Similarly, Ti-doped single crystals of the topological type-II Dirac semimetal NiTe$_2$ \cite{ferreira2021} exhibit experimental and band structure topology features reminiscent of the Cu$_x$ZrTe$_2$ compounds, which presents a superconducting state with T$_c\approx4.0$\,K \cite{de2018}. These findings underscore the importance of the intercalation-engineering mechanisms in TMDs, which bring about electronic signatures of great importance.

To the best of our knowledge, the coexistence of CDW and superconductivity in ZrTe$_2$ intercalated with Ni is reported in this work for the first time. High-quality single crystals were synthesized by isothermal chemical vapor transport (CVT). Electrical resistivity and specific heat data suggest the onset of a CDW ground state near 287\,K, followed by the onset of superconductivity with T$_c\approx4.1$\,K. The superconducting behavior suggests that the pairing charges come from two bands. First-principle electronic-structure calculations for ZrTe$_2$ and Ni-doped ZrTe$_2$ are provided, showing electronic and phononic dispersions that are consistent with the experimental findings.   

\section{Methods}
\label{sec:methods}

\subsection{Experimental Details}

The Ni$_{0.04}$ZrTe$_2$ single crystals were prepared by isothermal CVT. A precursor Ni$_{0.4}$Zr alloy was prepared initially by arc melting the stoichiometric amounts of high purity Ni and Zr in a Ti-gettered Ar atmosphere. The resulting material was ground together with Te for a Ni$_{0.4}$ZrTe$_2$ nominal composition, pelletized, sealed in a quartz tube under 0.4\,atm of UHP Ar, and reacted at 950\,$^\circ$C for 48\,hs. This material was quenched in water, an important step to avoid the formation ZrTe$_3$, reground, pelletized again, and treated again at 950\,$^\circ$C for 48\,hs. The Ni$_{0.4}$ZrTe$_2$ pellet was placed in a quartz tube containing $\approx 2.5$\,mg/cm$^3$ of iodine, which served as the transport agent, and sealed under vacuum. This tube was placed horizontally in a box furnace, the temperature was raised to 1000\,$^\circ$C and left there for 10 days. Given the off-stoichiometry composition of the charge, a gradient of the thermochemical potential forms around the pellet, creating the necessary conditions for chemical vapor transport growth. The tube was quenched again to avoid the formation of Te-rich phases and thin single crystals could be easily identified growing out of the pellet. The typical dimensions of the larger crystals were $ \approx 4 \times 4 \times 0.2$\,mm$^3$. 

The Ni-doped ZrTe$_2$ composition was determined by energy dispersive spectroscopy (EDS) and microwave plasma atomic emission spectroscopy (MP-AES), the later after dilution in HNO$_3$ and HCl. For MP-AES measurements, sample replicates, reagent blanks, and standard samples with known concentrations were included in each batch of analysis to ensure accuracy. A comparison of the different atom and ion emission lines of Zirconium, Tellurium and Nickel was carried out yielding a Ni$_{0.04}$ZrTe$_{1.89}$ composition. While the 1:2 ratio for Zr:Te is the same for all crystals, there is some variance in the concentration of Ni for crystals from the same batch, though the variance within each crystal is minimal. $\theta-2\theta$ X-ray diffraction (XRD) scans of the flat facets showed only the (00$l$) reflections of the ZrTe$_2$ structure, corresponding to the basal plane, which confirmed the phase purity and the single crystallinity. The Ni intercalation does not change the x-ray pattern in a significant way. A rocking curve centered on the (00$2$) reflection revealed a full-width spread at half-maximum within 0.05$^\circ$ (data not shown), suggestive of high crystallinity. The lattice parameter $c$ for Ni$_{0.04}$ZrTe$_2$ was determined from the $\theta-2\theta$ XRD scans, using the fitting tools from Powdercell \cite{kraus1996} and HighScore Plus (Panalytical) programs, yielding $c = 6.610 \pm 0.004$\,\AA. This value is consistent with the values of 6.63\,\AA \,\,reported for ZrTe$_2$ \cite{zhang2020} and 6.624\,\AA\,\, for Zr$_{1.16}$Te$_2$ \cite{kar2020}.

The measurements of electrical resistivity, specific heat and magnetization were carried out with options of the Quantum Design Physical Property Measurement System (PPMS-9), equipped with a 9\,T superconducting magnet. Resistivity measurements were performed using a 4-wire configuration, where 4 copper leads were attached to the sample using Ag paste. The typical contact resistance was $\approx 1.5\,\Omega$. For the measurements of specific heat, using a relaxation technique, a thin layer of N-grease was spread over the sapphire sample platform in order to provide thermal coupling between the sample, heater, and temperature sensor. The addenda was calibrated with an approximately 1 K spacing between points, in order to avoid convoluting the phase transition of the N-grease in the measurement \cite{bunting1969,quantum}. The magnetization measurements were performed using the vibrating sample magnetomer (VSM) of PPMS. For the magnetization measurements the samples were positioned with the $c$-axis perpendicular with the magnetic field.

\begin{figure*}
	\centering
	\includegraphics[width=.6\textwidth]{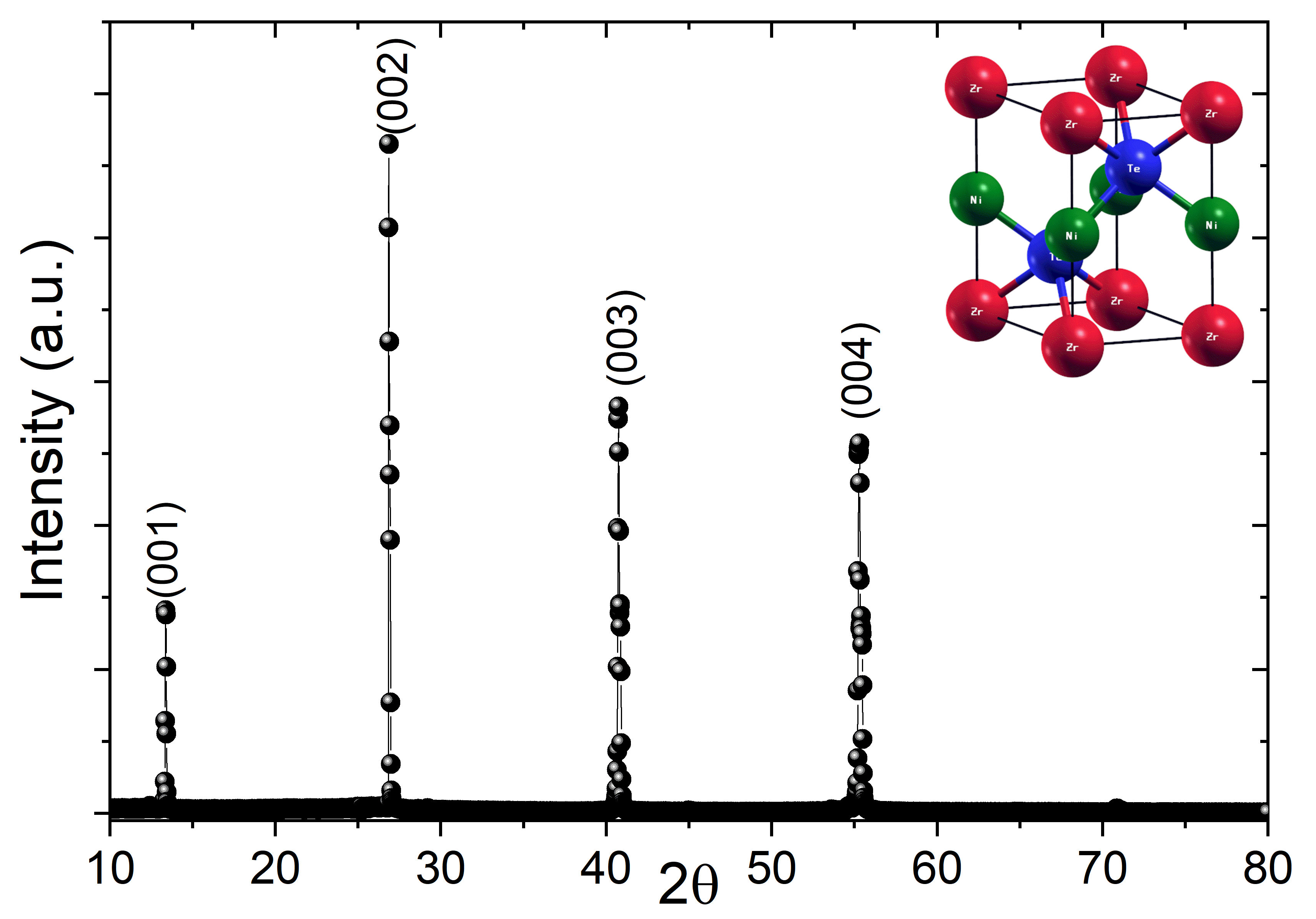}
	\caption{Powder x-ray diffraction of Ni$_{0.04}$ZrTe$_{1.89}$. The inset shows the unit cell of NiZrTe$_2$ (Ni: green; Zr: red; Te: blue).}
	\label{fig:DRX}
\end{figure*}

\subsection{DFT calculations}
\label{sec:computational}

First-principles electronic-structure calculations were conducted within the Kohn-Sham scheme of the Density Functional Theory (DFT) \cite{hohenberg1964, kohn1965} using ultrasoft pseudopotentials \cite{dal2014}, as implemented in Quantum \textsc{espresso} \cite{giannozzi2009, giannozzi2017}, and some auxiliary post-processing tools \cite{kokalj1999, kawamura2019}. The calculations were performed using the generalized gradient approximation (GGA) by Perdew-Burke-Ernzerhof (PBE) \cite{perdew1996} for the exchange and correlation (XC) effects. All numerical and structural parameters were converged and  optimized to guarantee a ground state convergence of 10$^{-6}$\,Ry in total energy and 10$^{-4}$ Ry/a$_0$ (a$_0 \approx 0.529$\,\AA) in total force acting on nuclei. Phonon dispersions were obtained by Density Function Pertubation Theory \cite{giannozzi2009, giannozzi2017} using van der Waals non-local corrections \cite{thonhauser2007,langreth2009,sabatini2012,hamada2014,thonhauser2015,berland2015}. The dynamical matrix was calculated using a $3\times3\times2$ $q$-point grid and a Marzari-Vanderbilt-DeVita-Payne cold smearing value of 0.04\,Ry \cite{marzari1999}. In order to simulate different compositions of Ni$_x$ZrTe$_2$, we used fully-relaxed $2 \times 2 \times 1$ supercells, thus neglecting possible disorder effects \cite{okhotnikov2016}. 

\section{Results and Discussion}
\label{sec:results}

\subsection{Resistivity}

The temperature dependence of the electrical resistivity $\rho(T)$ for Ni$_{0.04}$ZrTe$_2$ in the $ab$-plane is shown in Fig. \ref{fig:rho}(a). Superimposed to the metallic behavior, two features can be clearly identified. First, is the occurrence of a superconducting transition with onset at T$_c\approx4.1$\,K. This value of T$_c$ was taken from the temperature where an extrapolation of $\rho(T)$ from the normal phase and a fit to the $\rho(T)$ data below T$_c$ start to diverge. The $\rho(T) = 0$ value is reached near 2.7\,K. Secondly, a hump centered near 287\,K,  taken from the discontinuity in $d\rho/dT$, is suggestive of the formation of  charge density waves \cite{berthier1976,craven1977,yang2014,zhu2015,kolincio2017}. In fact, the behavior of $\rho(T)$ for Ni$_{0.04}$ZrTe$_2$ reflecting the coexistence of the superconducting and CDW ground states is reminiscent, for example, of the behavior found in the low-dimensional compounds such as ErTe$_3$ and HoTe$_3$ \cite{pfuner2010}, 2H-Pd$_x$TaSe$_2$ \cite{bhoi2016} and HfTe$_3$ \cite{denholme2017}.

\subsection{Specific heat}

\begin{figure*}[t]
	\centering
	\subfloat[][]{\includegraphics[width=.33\textwidth]{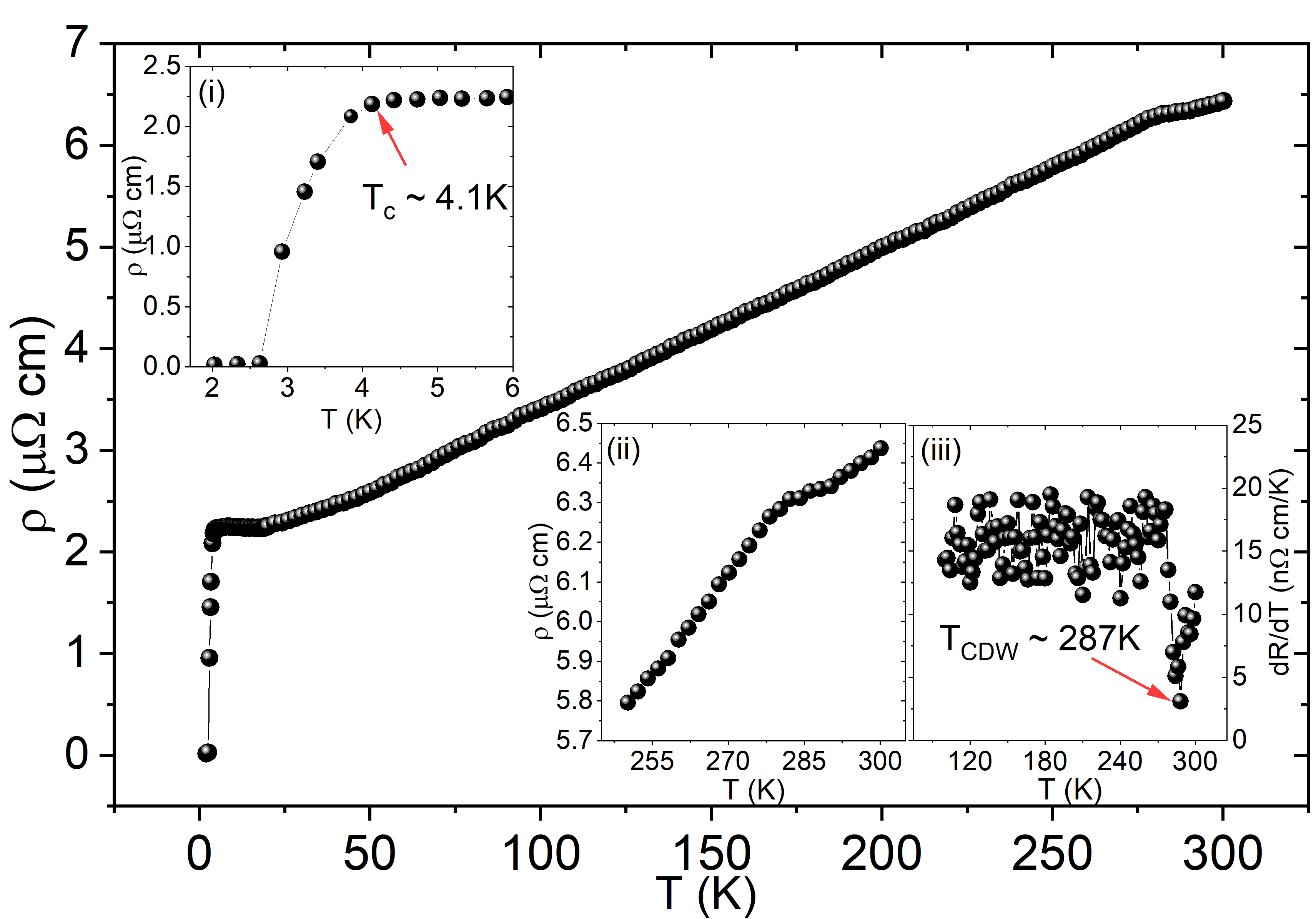}}
	\subfloat[][]{\includegraphics[width=.33\textwidth]{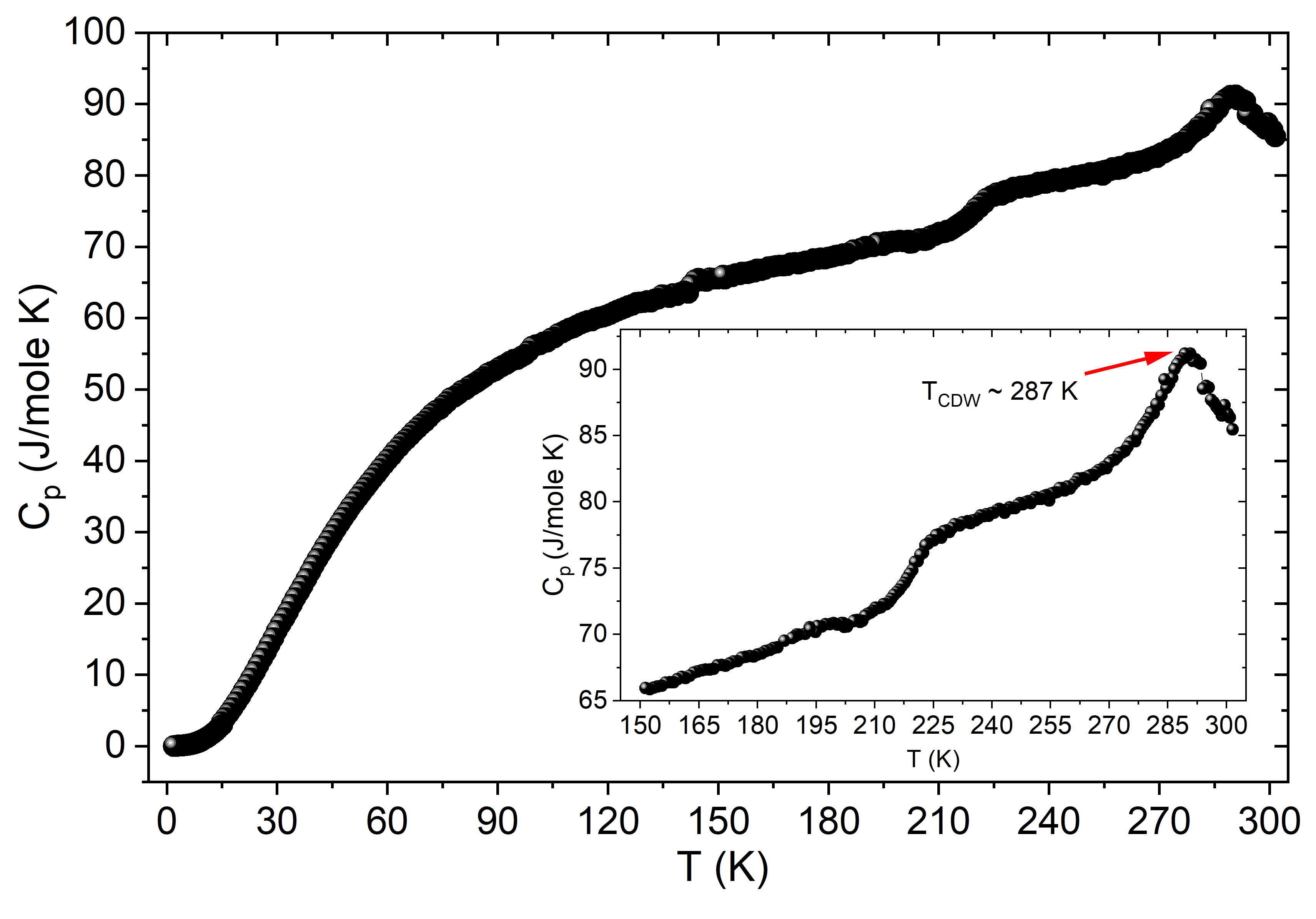}}
	\subfloat[][]{\includegraphics[width=.29\textwidth]{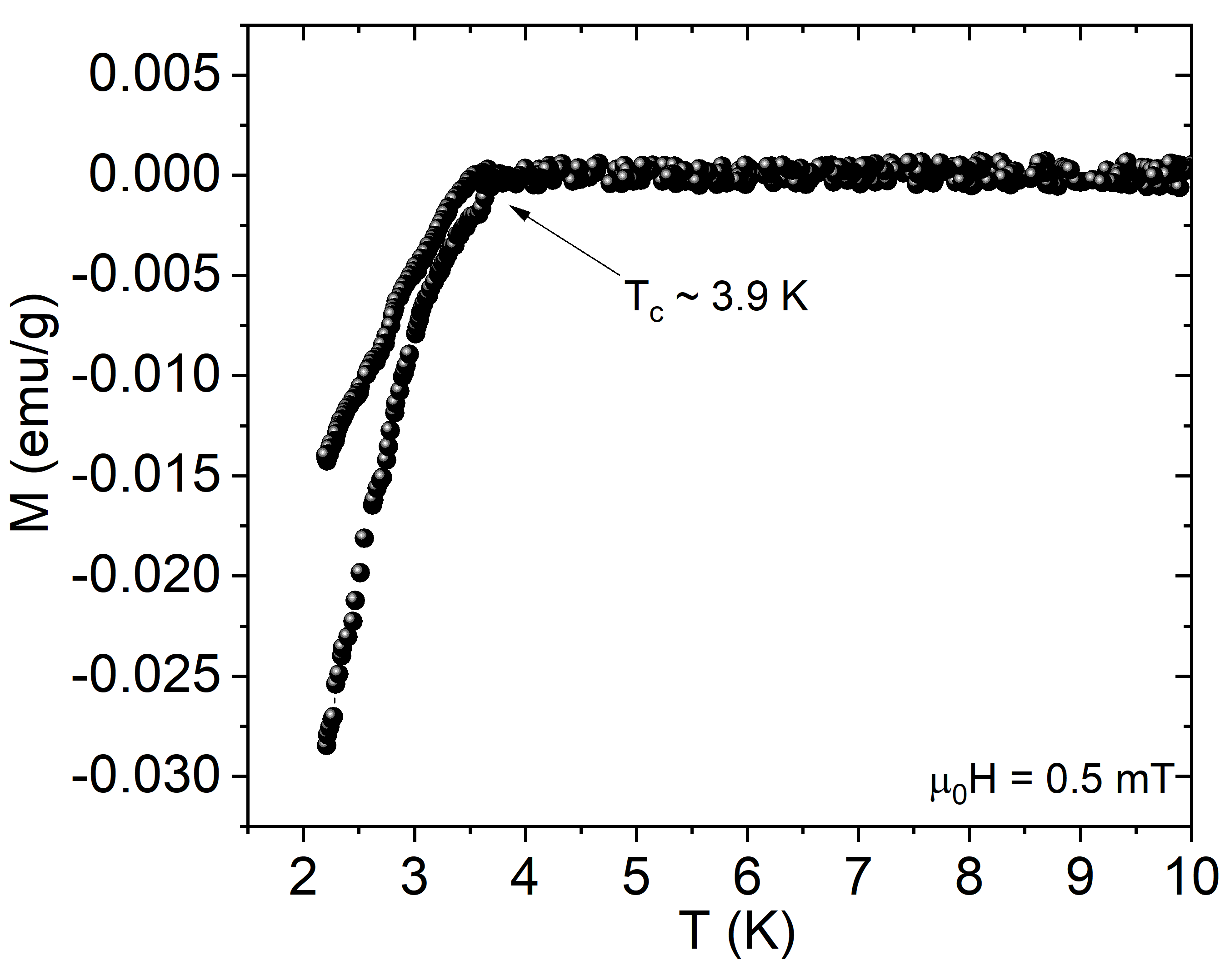}}
	\caption{(a) Resistivity dependence with temperature for Ni$_{0.04}$ZrTe$_2$ zero magnetic field. The insets show the $\rho(T)$ curve in (i) low-temperature regime and (ii) high-temperature regime. The inset (iii) presents the $d\rho/dT$ curve, evidencing the CDW transition at $T_{CDW} \approx 287$\,K. (b) Temperature dependence of the specific heat for Ni$_{0.04}$ZrTe$_2$. The inset depicts the CDW anomaly at high temperatures. (c) Low-field ZFC and FC regime of the magnetization in function of the temperature at 0.5\,mT.}
	\label{fig:rho}
\end{figure*}

The temperature dependence of the specific heat for Ni$_{0.04}$ZrTe$_2$ is shown in Fig. \ref{fig:rho}(b). A fit of the C$_p$/T vs T$^2$ data at low temperatures to $\gamma + \beta T^2$ yields values for the Sommerfeld coefficient $\gamma$ and the phonon term $\beta$ equal to 4.79\,mJ/mole\,K$^2$ and 0.99\,mJ/mole\,K$^4$, respectively, the latter corresponding to a Debye temperature $\Theta_D \approx 380$\,K. In contrast to the $\rho(T)$ data, there is no clear feature suggestive of a superconducting transition in C$p$, which raises the question of whether superconductivity in Ni$_{0.04}$ZrTe$_2$ is a bulk phenomenon.

However, the C$_p$ vs T data of Fig. \ref{fig:rho}(b) show a clear feature with onset near 287\,K, which correlates well in temperature with the feature in $\rho(T)$, both consistent with the onset of a CDW order. It is plausible that a second feature in the $C_p(T)$ data centered near 215 K is also due to a CDW transition even though no corresponding anomaly could be found in the $\rho(T)$ data. Two CDW transitions at relatively closed temperatures have already been identified in trichalcogenides HoTe$_3$ (265 and 288\,K) and ErTe$_3$ (110 and 157\,K) \cite{pfuner2010}. In order to probe this point further, an estimate of the portion of the Fermi surface gapped at each of the two $C_p(T)$ features in Ni$_{0.04}$ZrTe$_2$ can be made by evaluating the change in entropy associated with each one. A simplified analysis of the $C_p/T$ vs $T$ data from the onset to completion of each of the two CDW features, relative to the background, yielded gaps of $\approx$ 3\,\% and $\approx$ 10\,\% of the Fermi surface, for the feature at 215 and 287\,K, respectively. The overall trend of $\rho(T)$ is still metallic. While the 10\,\% gapping of the Fermi surface corresponds to a noticeable upturn in $\rho$(T) near 287\,K, possibly the 3\,\% gapping at 215 K is not sufficiently large to be unequivocally identified in the $\rho(T)$ data.

Given a phase transformation near 290\,K in N-grease \cite{bunting1969,quantum}, which we use as the thermally coupling agent between the sample and the calorimeter platform, it is important to rule out experimental artifacts as the cause for the 287\,K feature in C$_p$ vs T. This was accomplished by spacing the calorimeter’s addenda calibration closely, in this case $\approx 0.667$\,K in the 2--300\,K range. We also confirmed that the $C_p(T)$ features remained unchanged when H-grease is used as the thermal coupling agent (data not shown). It is worth mentioning that it was reported an interface and/or Zr intercalation induced CDW order in ZrTe$_2$ thin films prepared on graphitized SiC(0001) substrates \cite{ren2021}. This finding reinforces the ideia of a hidden CDW order in ZrTe$_2$, possibly reminiscent of TiSe$_2$ \cite{zunger1978, fang1997, bianco2015, das2015}.

\subsection{Magnetization}

\begin{figure*}[t]
	\centering
	\subfloat[][]{\includegraphics[width=.45\linewidth]{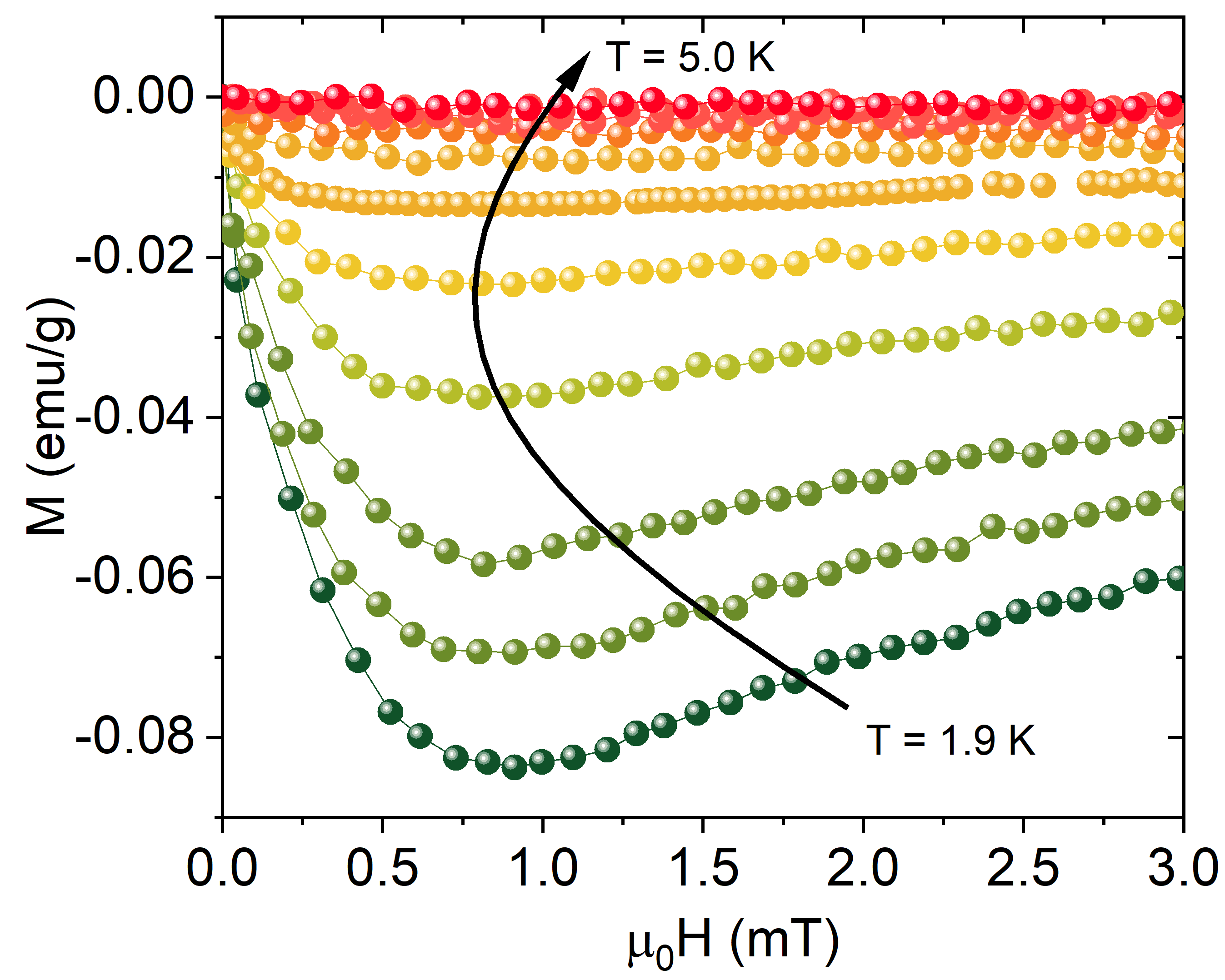}}
	\subfloat[][]{\includegraphics[width=.45\linewidth]{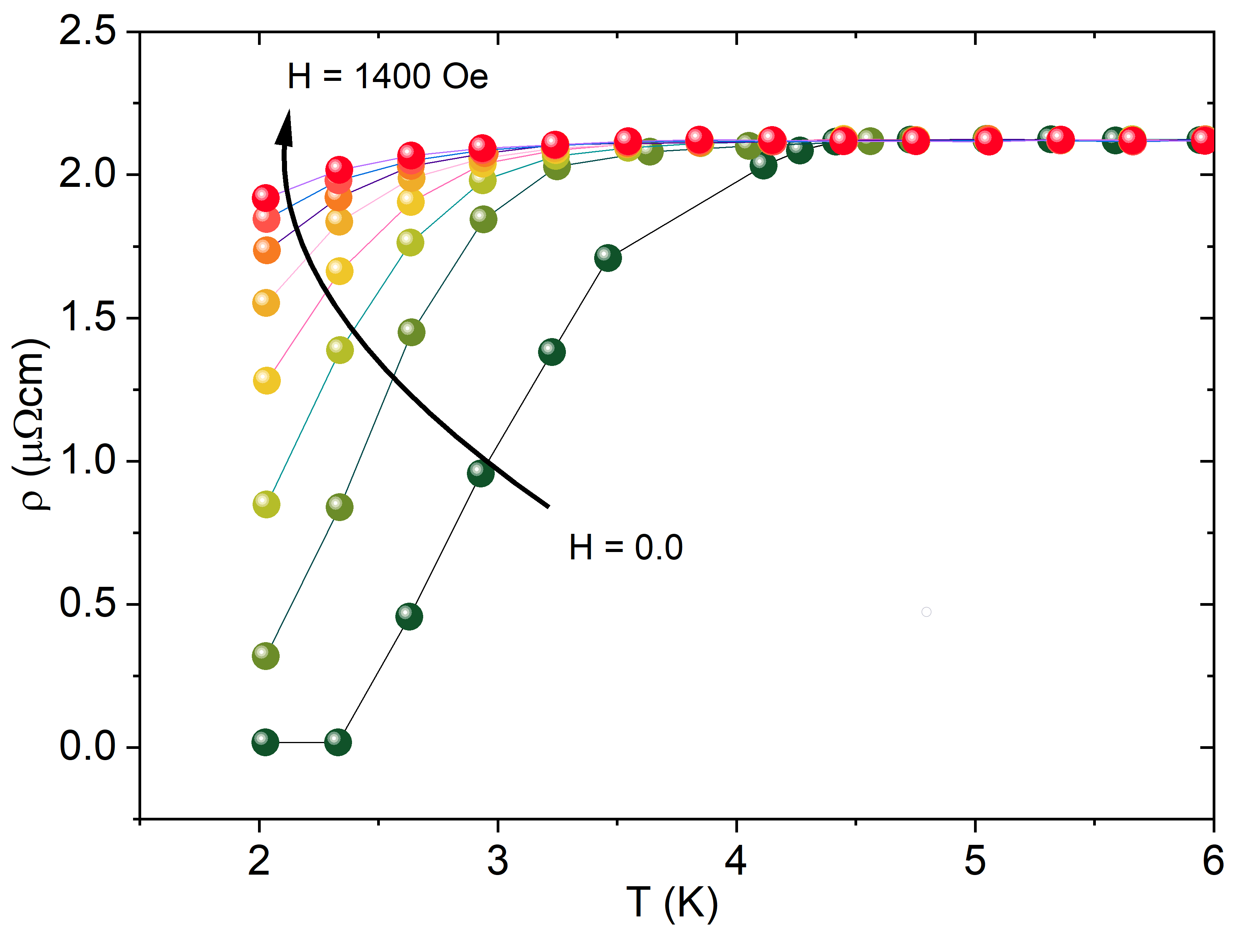}}\\
	\subfloat[][]{\includegraphics[width=.47\linewidth]{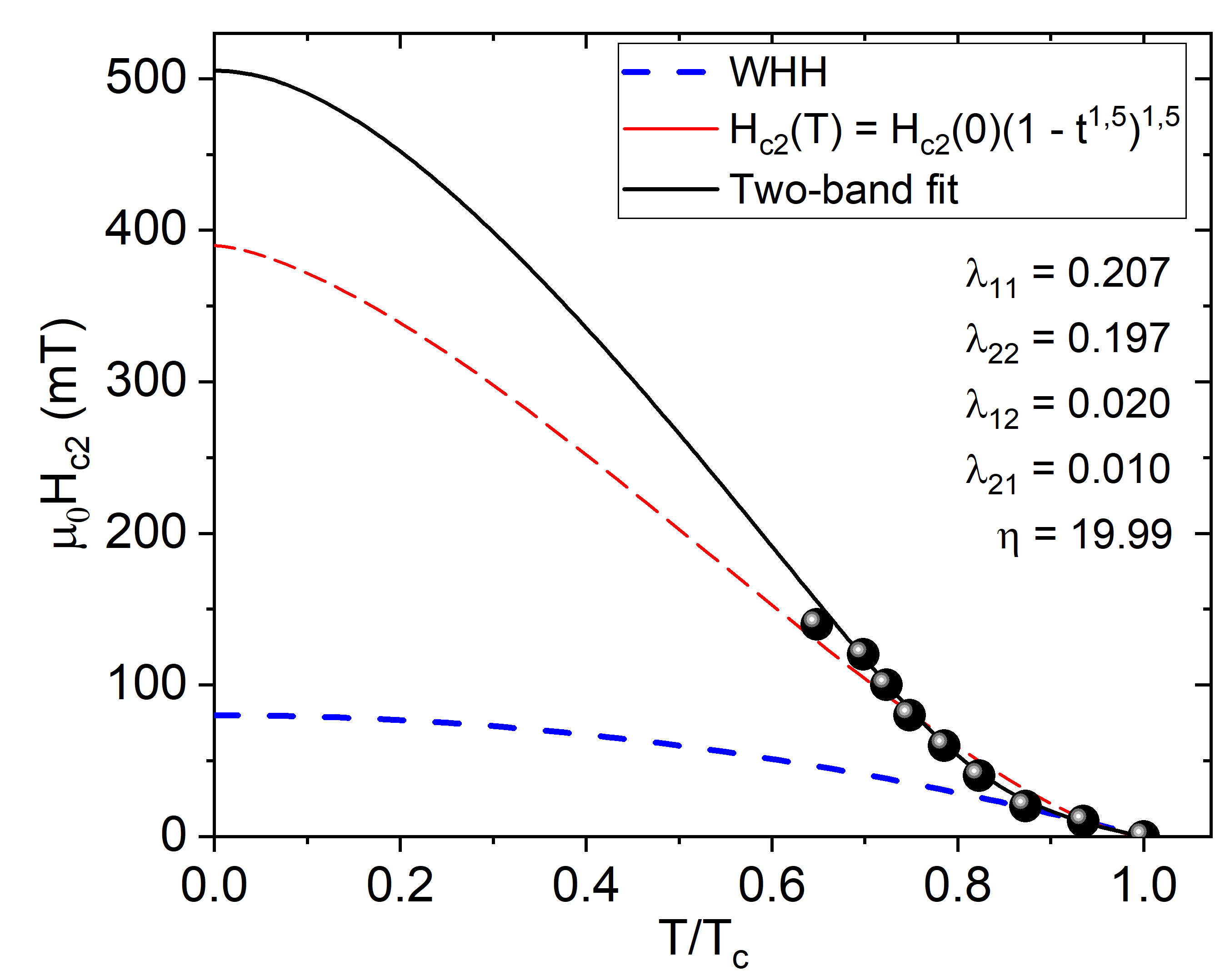}}
	\subfloat[][]{\includegraphics[width=.45\linewidth]{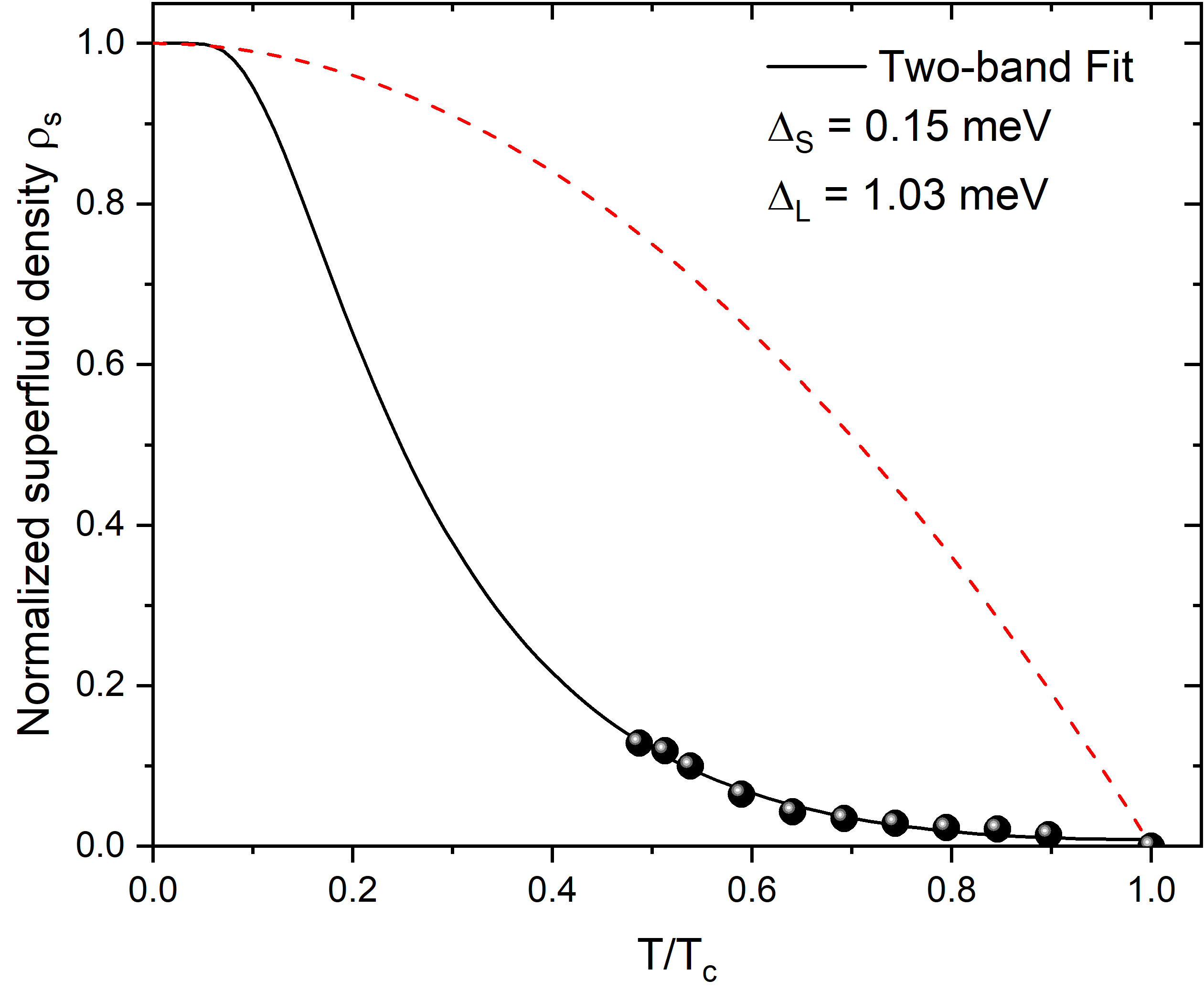}}\\
	\caption{(a) Magnetization dependence with the applied magnetic field for Ni$_{0.04}$ZrTe$_2$. (b) Magnetoresistance with applied magnetic field. (c) Upper critical field as a function of reduced temperature. (d) Normalized superfluid density as a function of reduced temperature at zero magnetic field. The red dashed line shows the expected behaviour for an isotropic single-gap pairing mechanism.}
	\label{fig:Hc}
\end{figure*}

To determine the lower and upper critical fields, H$_{c_1}$ and H$_{c_2}$, respectively, we carried out isothermal magnetization curve measurements from 1.9 to 5.0\,K, and measurements of $\rho(T)$ near T$_c$ in fields up to 1400\,Oe, as shown in Fig. \ref{fig:Hc}. These data were not corrected for the demagnetization factor. In spite of the specific heat data lacking a clear anomaly due to superconductivity, the magnetization curves below T$_c$ [Fig. \ref{fig:Hc}(a)] and the drive of the resistivity transitions at T$_c$ to lower temperatures with magnetic field
[Fig. \ref{fig:Hc}(b)] are consistent with superconductivity. Indeed, the plot of low field ZFC and FC $M(T)$ [Fig.\ref{fig:rho}(c)] shows coherently that the superconducting transition is about 3.9\,K, with a superconducting fraction volume of approximately 62\,\%.

The temperature dependence of H$_{c_2}$ for Ni$_{0.04}$ZrTe$_2$ is shown in Fig. \ref{fig:Hc}(c). The H$_{c_2}$ values were taken from onset of the transitions in $\rho(T)$, i.e., the first instances in which a separation between the extrapolated values of the normal state. The behavior of H$_{c_2}(T)$ in the dirty limit, within the framework of the single-band BCS theory, was modeled by Werthamer-Helfand-Hohenberg (WHH) model \cite{WHH}. Estimating the derivative $dH_{c_2}/dT|_{T=T_c}$ from the data in Fig. \ref{fig:Hc}(c) and using the expression  
\begin{align}
	\mu_0 H_{c_2}(0) = -0.693T_c\dfrac{dH_{c_2}}{dT}\mid_{T=T_c}
\end{align}
the fit yields the blue line in Fig. \ref{fig:Hc}(c), which obviously underestimates the values of H$_{c_2}$. 
The upturn in the H$_{c_2}(T)$ data, which starts to become more prominent near T/T$_c \approx 0.85$, is in sharp contrast with the quadratic behavior of  the WHH theory. Actually, the positive curvature of $\mu_0 H_{c_2}$ below T$_c$ is an ubiquitous feature in multiband superconductivity \cite{lyard2002, hunte2008, lei2012, li2018, santos2018, shang2019, xu2019, majumdar2020}.

Ignoring the interband impurity scattering channel and taking the orbital pair-breaking effect, we can fit the H$_{c_2}$ within a two-band model derived from quasi-classic Usadel equations \cite{gurevich2003}, given by
\begin{align}
 0 = &a_0\left[\ln t + U(h)\right]\left[\ln t + U(\eta h)\right] \nonumber \\ 
 &+ a_1\left[\ln t + U(h)\right] + a_2\left[\ln t + U(\eta h)\right],  
\label{eq:hc2}
\end{align}
where $a_0$, $a_1$ and $a_2$ are defined by the intraband ($\lambda_{11}$ and $\lambda_{22}$) and interband ($\lambda_{12}$ and $\lambda_{21}$) coupling strength between bands 1 and 2, $h = H_{c2}D_1/2\phi_0T$, $\eta = D_2/D_1$, and $U(x) = \psi(x+1/2) - \psi(1/2)$. The $\psi(x)$ is the digamma function, $D_1$ and $D_2$ are the intraband electronic diffusivities of bands 1 and 2 in the normal-state, and $\phi_0$ is the magnetic flux quantum. The diffusivity ratio $\eta$ determines the curvature of H$_{c2}(T)$. The Ginzburg-Landau parameter is considered temperature independent.

As seen in Fig. \ref{fig:Hc}(c), the H$_{c_2}(T)$ data can be fit very well with this two-band model. The upper critical field at $T = 0$ yielded by Equation \ref{eq:hc2} is $H_{c_2}(0) = 5.05$\,kOe. The resultant diffusivity and coupling parameters are $\eta = 19.99$, $\lambda_{11} = 0.207$, $\lambda_{22} = 0.197$, $\lambda_{12} = 0.02$ and $\lambda_{21} = 0.01$. The value found for the diffusivity ratio reflects a large difference in the electron mobility of the Fermi surface sheets involved in the superconductivity, originating the positive curvature that deviates from the traditional WHH-like single-band behavior, and reflects the suppression of the small superconducting gap with magnetic field. Moreover, values of $\lambda_{ij}$ from the fit suggest that the intraband coupling is one order of magnitude higher than the interband scattering, also favoring the emergence of the two-gap superconductivity \cite{silaev2011,silaev2012,cavalcanti2020}.

The values of H$_{c_1}$ were extracted from the $M(H)$ isotherms of Fig. \ref{fig:Hc}(a) by using  $\Delta M = 10^{-3}$\,emu/g criterion, i.e. the values of H$_{c_1}$ at each temperature were taken when the difference between the data and the extrapolated linear regions reached $10^{-3}$\,emu/g. The temperature dependence of the superfluid density $\rho_s(T)$ for Ni$_{0.04}$ZrTe$_2$, normalized to its zero temperature value, is shown in Fig. \ref{fig:Hc}d. In the framework of the London approximation, $\rho_s(T) = H_{c_1}(T)/H_{c_1}(0)$. The lower critical field behavior of Fig. \ref{fig:Hc}(d) diverges quite significantly from the quadratic behavior expected from the single-band model (dashed line). A robust upturn in H$_{c_1}$ becomes quite noticeable near $T/Tc\approx0.75$. Anomalous upturns in H$_{c_1}$ of this type have consistently been found in two-band superconductors \cite{angst2002, ren2008}.

For an isotropic two-gap superconductor in the Meissner state, the normalized superfluid density at low temperatures can be expressed as \cite{kim2002}
\begin{align}
\rho_s(T) &= 1 - c\left(\dfrac{2\pi\Delta_S(0)}{k_BT}\right)^{1/2}e^{-\dfrac{\Delta_S(0)}{k_BT}}\nonumber \\ 
&-\left(1-c\right)\left(\dfrac{2\pi\Delta_L(0)}{k_BT}\right)^{1/2}e^{-\dfrac{\Delta_L(0)}{k_BT}},
\label{eq:superfluid}
\end{align}
where $\Delta_S$ and $\Delta_L$ are the small and large superconducting gaps, respectively, and $c$ is the superconducting weight parameter of $\Delta_S$. The solid line in Fig. \ref{fig:Hc}(d) shows the fit to the superfluid density using Equation \ref{eq:superfluid}. The good fit is consistent with the existence of two superconducting gaps at the Fermi surface.  The estimated parameters for Ni$_{0.04}$ZrTe$_2$ are $\Delta_S(0) = 0.15$\,meV, $\Delta_L(0) = 1.03$\,meV, $c = 0.09$, and H$_{c_1}(0) = 35$\,Oe. Alternatively, for the case of a single anisotropic superconducting gap, the $c$ parameter would be 0 or 1. However, given that both $\Delta_S$ and $\Delta_L$ terms have non-zero values, the possibility that the anomalous upturn in $H_{c_1}(T)$ stems from a single anisotropic order parameter can be ruled out. 

The dashed line in Fig. \ref{fig:Hc}d shows the expected behavior for $\rho_s(T)$ if the Ni$_{0.04}$ZrTe$_2$ superconductivity stemmed from a single-gap, considering an isotropic Fermi velocity and s-wave pairing symmetry \cite{chandrasekhar1993, carrington2003,prozorov2006}. The strong discrepancy between the two models excludes the single isotropic s-wave gap scenario. In light of the impossibility of matching the $H_{c_1}(T)$ and $H_{c_2}(T)$ behaviors to a single-gap model, and the success of the two-gap model to fit the data, the latter gains credence, without the need to invoke unconventional pairing symmetries or strong anisotropic effects for the Fermi surface.

\subsection{CDW and superconductivity interplay: first-principles calculations} 

\begin{figure*}[t]
    \centering
	\includegraphics[width=.6\textwidth]{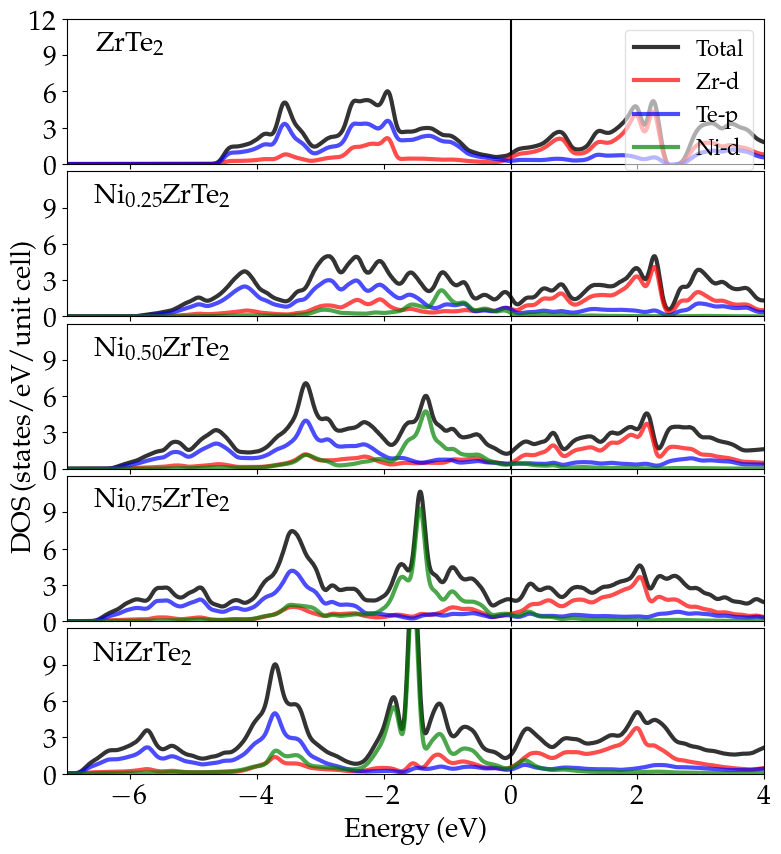}
	\caption{Partial density of states for different Ni-doped Ni$_x$ZrTe$_2$ compounds.}
	\label{fig:DOS}
\end{figure*}

Given the strong possibility of multiband superconductivity and CDW coexistence/competition in Ni-doped ZrTe2, as yielded by the resistivity, magnetization and heat capacity data, we carried out first-principles calculations of the electronic and phononic structures in order to gain insights on the microscopic mechanisms in this system. The partial density of states for Ni$_x$ZrTe$_2$ ($x = 0.00, 0.25, 0.50, 0.75, 1.00$) is presented in Fig. \ref{fig:DOS}. The total density of states at the Fermi level ($N_{E_F}$) including spin-orbit coupling effects of pure ZrTe$_2$ is 0.985\,states/eV. Of these carriers, 62\% are derived from Zr-3d manifold and 34\% corresponds to Te-5p orbitals. As the Ni intercalating content increases, the Ni-3d states create a localized band around $-1.8$\,eV, where a narrow peak in the density of states appears, which indicates the formation of localized, near-flat bands. A second, less intense peak is also found in the region close to $-4$\,eV. However, in the vicinity of the Fermi level, only a slight hybridization of the Te-5p and Ni-3d wave-functions starts to develop. At $x=0.25$, for instance, the total DOS increases to 1.62\,states/eV, thus favoring the emergence of spontaneous symmetry breaking mechanisms such as superconductivity and other electronic instabilities. At the same time the contribution of the Te-p manifold at the Fermi surface reaches 64\,\% of the total DOS. Therefore, the main role of the Ni ions as they populate the van der Waals gap is the stabilization of the transferred charge to the Te sites, gradually enhancing the electronic correlation at the Fermi surface consistently with the doping level. In fact, this mechanism is similar to the formation of strong Cu-Se bonds in Cu$_x$TiSe$_2$ \cite{jishi2008}.

From the estimated total density of states at the Fermi level for Ni$_{0.04}$ZrTe$_2$ we can evaluate the superconducting critical temperature using McMillan's equation \cite{mcmillan1968},
\begin{equation}
	T_c = \dfrac{\Theta_D}{1.45}\exp\left[-\frac{1.04(1+\lambda)}{\lambda - \mu^{*}(1 + 0.62\lambda)}\right],
	\label{eq:mcmillan}
\end{equation}  
where $\Theta_D$ is the Debye temperature, $\lambda$ is the electron-phonon coupling constant, and $\mu^{*}$ is the Coulomb pseudopotential, which measures the strength of the electron-electron Coulomb repulsion \cite{mcmillan1965}. The electron-phonon coupling constant and Coulomb pseudopotential, respectively, are estimated to be 0.70 and 0.17 according to $\lambda \approx (\gamma_{exp}/\gamma_{calc}) - 1$ and $\mu^* \approx \{0.26N(E_F)/[1+N(E_F)]\}$ \cite{ferreira2018, bennemann1972}, and the estimated critical temperature from Eq. \ref{eq:mcmillan} is T$_c \approx 5.28$\,K, close to the 4.1\,K value from the $\rho(T)$ data. The proximity of the measured and estimated values is consistent with the electron-phonon interactions being the driving mechanism for superconductivity. 

\begin{figure*}
	\centering
	\subfloat[][]{\includegraphics[width=.45\linewidth]{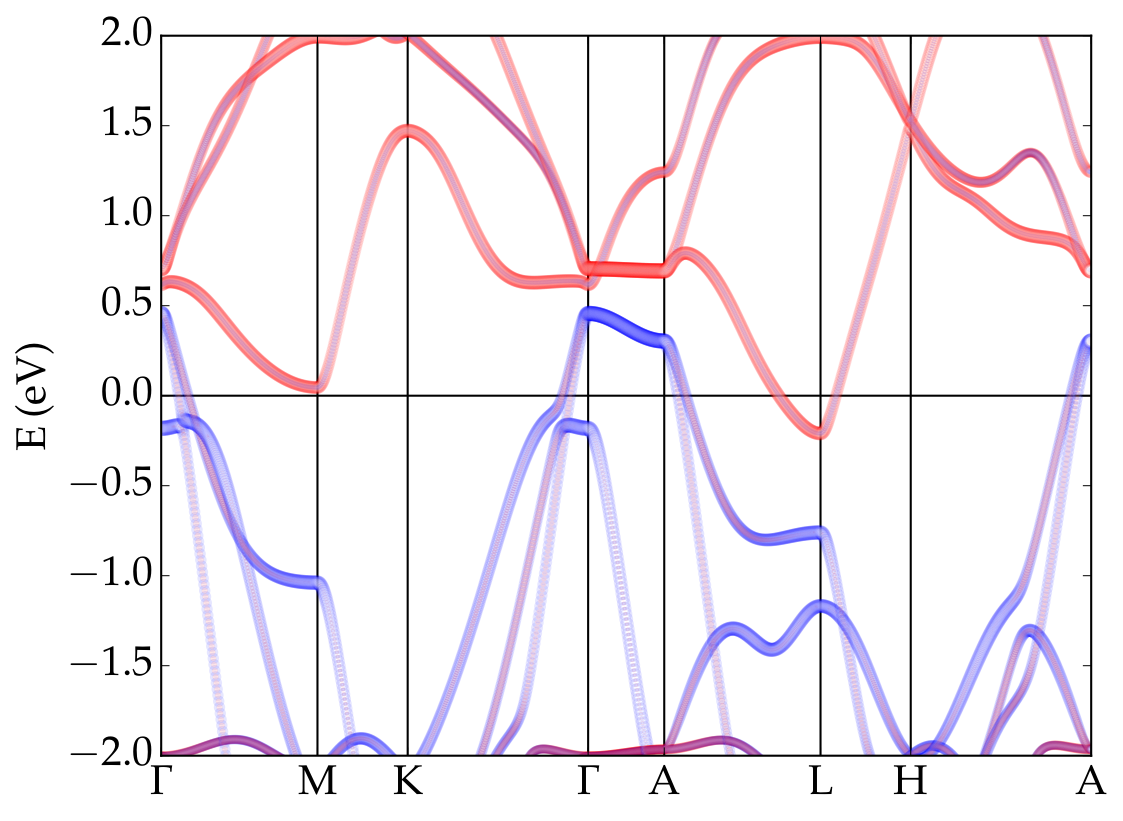}}
	\subfloat[][]{\includegraphics[width=.45\linewidth]{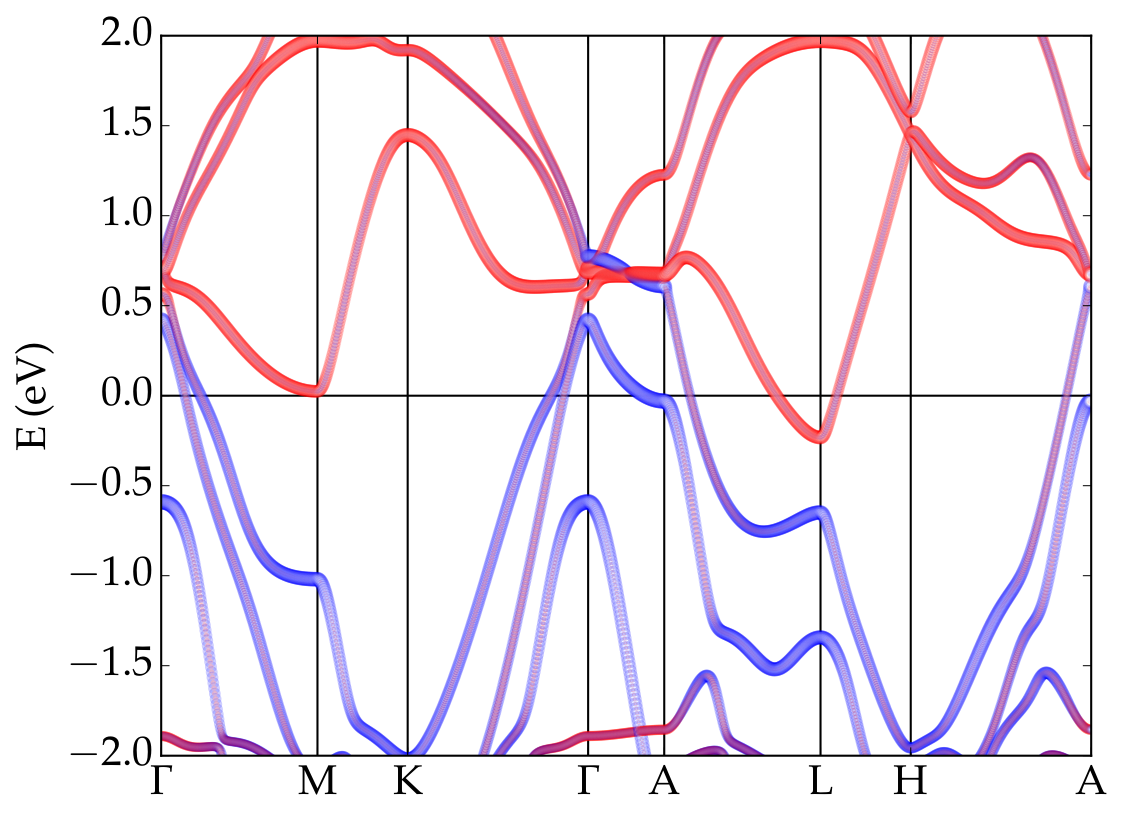}}\\
	\subfloat[][]{\includegraphics[width=.24\linewidth]{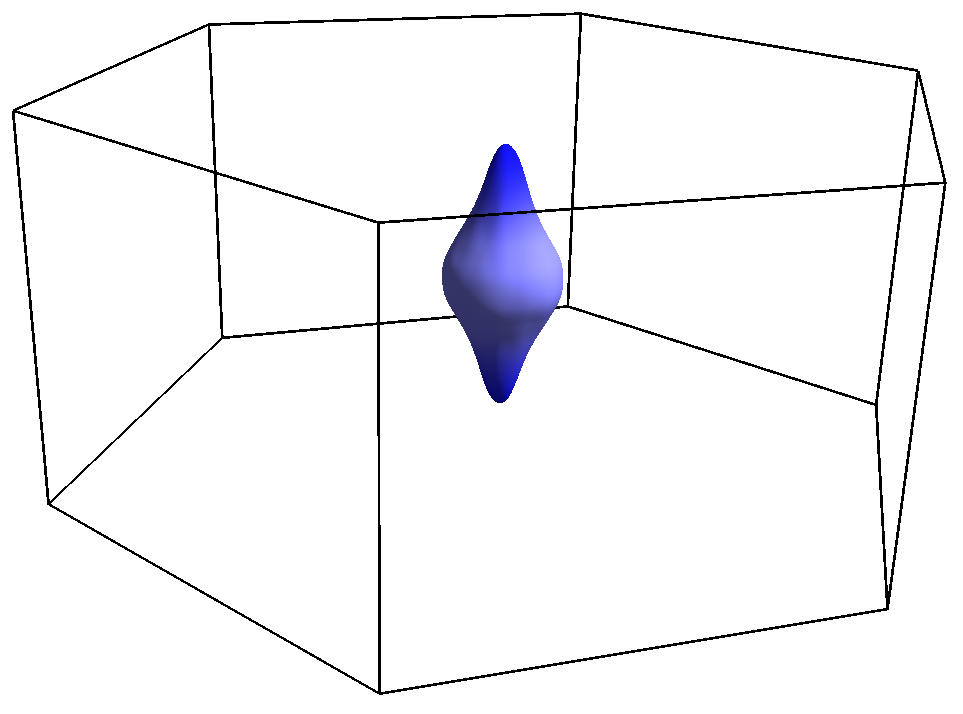}}
	\subfloat[][]{\includegraphics[width=.24\linewidth]{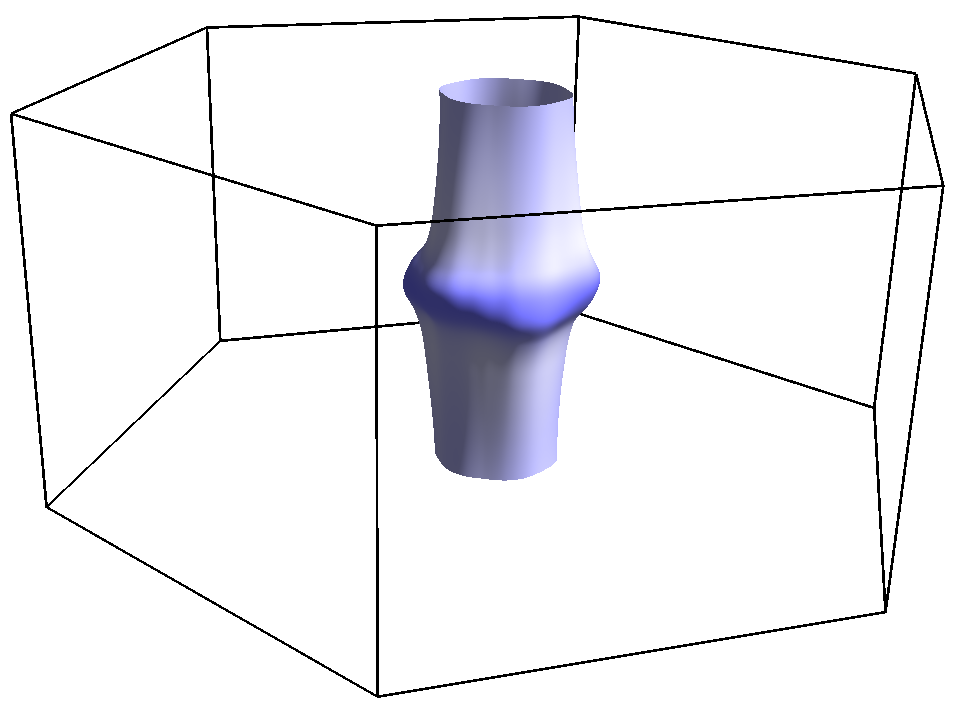}}
	\subfloat[][]{\includegraphics[width=.24\linewidth]{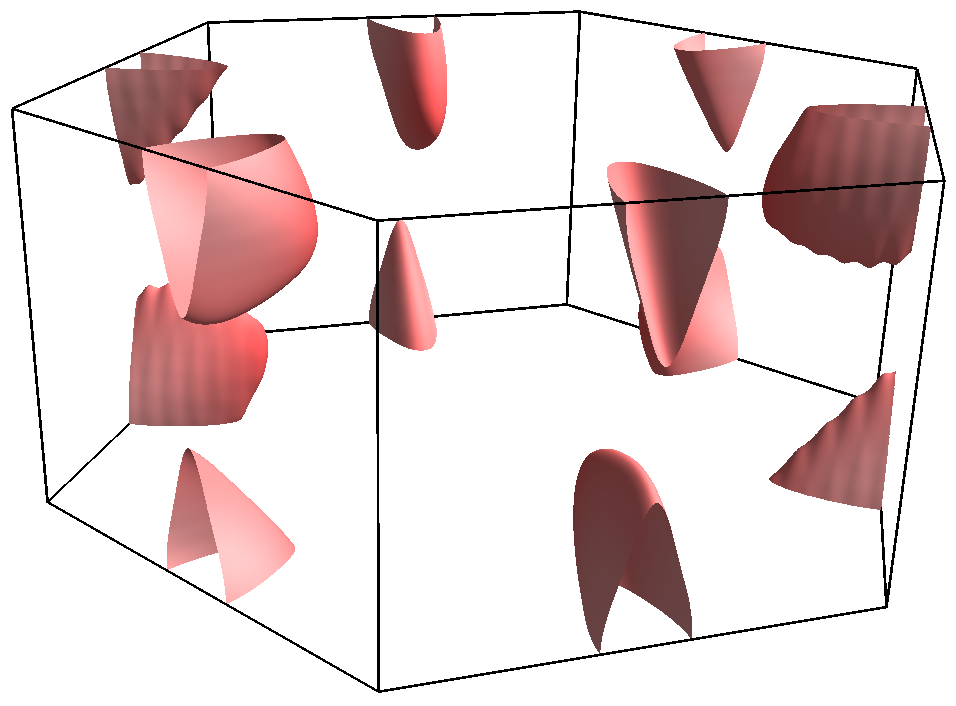}}\\
	\includegraphics[width=.36\linewidth]{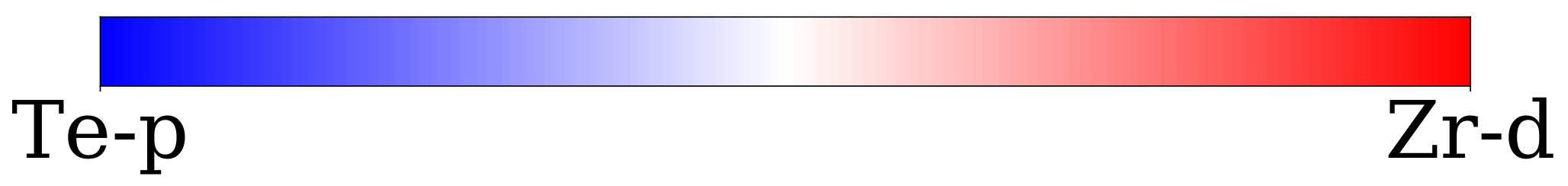}\\
	\subfloat[][]{\includegraphics[width=.24\linewidth]{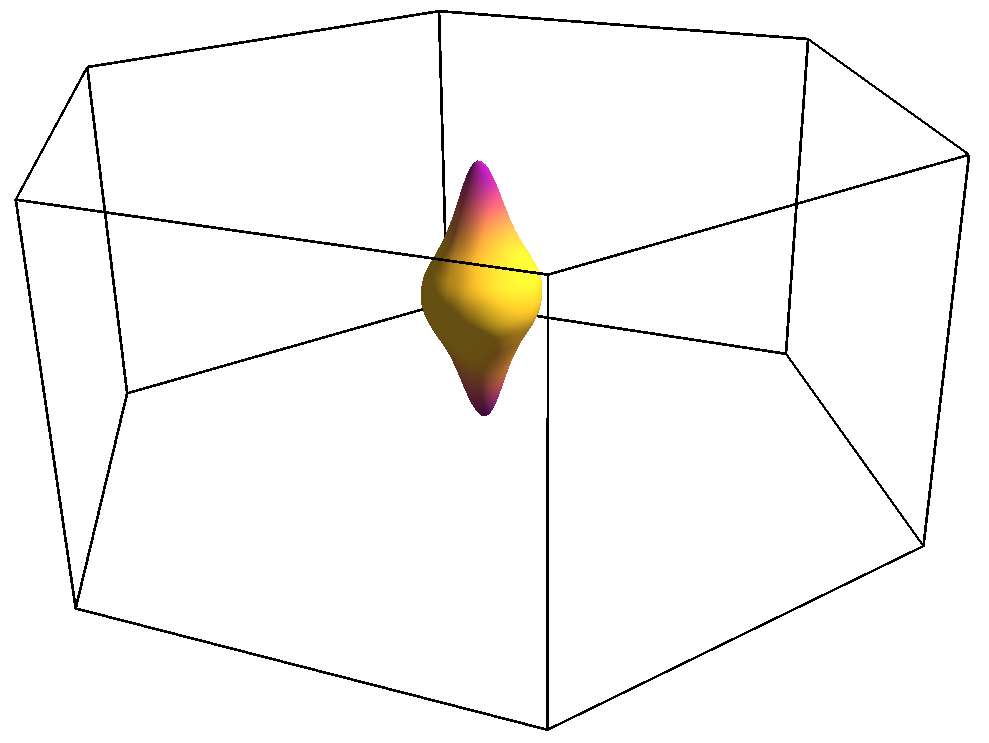}}
	\subfloat[][]{\includegraphics[width=.24\linewidth]{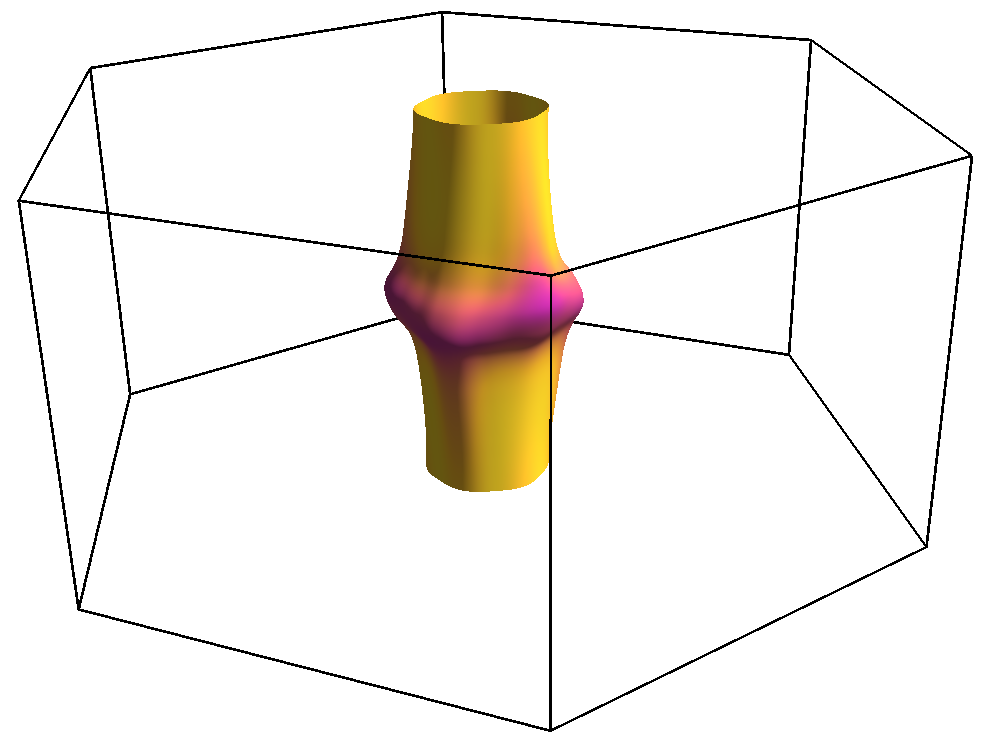}}
	\subfloat[][]{\includegraphics[width=.24\linewidth]{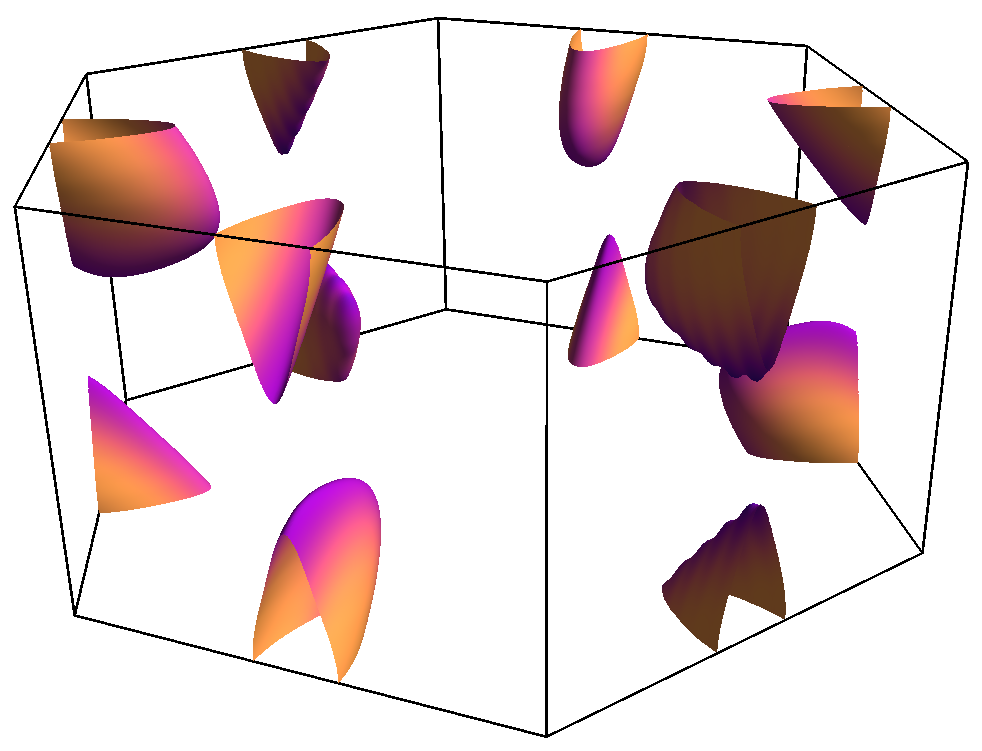}}\\
	\includegraphics[width=.35\linewidth]{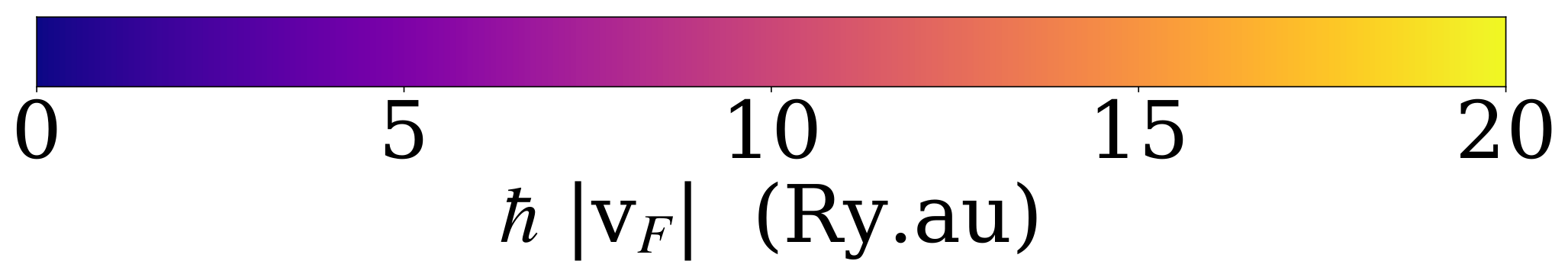}
	\caption{Electronic band structure (a) without and (b) with spin-orbit coupling effects along high-symmetry points at the first Brillouin zone projected onto Te-$p$ (blue) and Zr-$d$ (red) orbitals. (c)-(e) the three independent sheets of the ZrTe$_2$ Fermi surface. The color map shows the contribution of Te-5p (blue) and Ni-3d (red) manifold to the electronic wave function. (f)-(h) Fermi surface projected onto the Fermi velocity.}
	\label{fig:band}
\end{figure*} 

In particular, it is reasonable to assume that the electronic ground state of Ni$_{0.04}$ZrTe$_2$ corresponds to the energy-momentum dispersion of ZrTe$_2$, with a small displacement of chemical potential due to Ni-$d$ hybridization with Te-$p$ orbitals. For comparison, the electronic band-structure without and with spin-orbit effects and the Fermi surface topography of ZrTe$_2$ is presented in Fig. \ref{fig:band}. The calculated electronic ground state  is in excellent agreement with recent spectral density maps obtained by high-resolution ARPES measurements \cite{kar2020}. The typical semimetal behavior of $\rho(T)$ data for pure ZrTe2 \cite{aoki1996} is consistent with the 3 distinct bands crossing the Fermi level. Two hole-like bands ($\alpha_1$ and $\alpha_2$) cross the Fermi level around $\Gamma$ and A high-symmetry points, giving rise to quasi-cylindrical pockets along the out-of-plane direction in the Fermi surface, and an additional electron-like band ($\beta_1$) can be found around the L direction in the first Brillouin zone, which compose multiple disconnected pockets with hexagonal symmetry. We can observe a clear predominance of Zr-d orbitals in the conduction, electron band ($\beta_1$ pocket) and a majority contribution of Te-p orbitals in valence, hole bands ($\alpha_1$ and $\alpha_2$ pockets). The disconnected multiband nature with very distinct orbital projections, Fermi velocities, and nesting vectors support two-band superconductivity conjecture in this system \cite{xu2019, floris2007, singh2013, singh2015, ferreira2018, bhattacharyya2020, zhao2020, de2021}, consistently with our experimental findings. In fact, the different band Fermi velocities of ZrTe$_2$  favors the emergence of such unconventional superconducting magnetic properties due to the presence of multiple condensates with dissimilar competing characteristic lenghts \cite{silaev2011,silaev2012,chen2020}.

It is interesting to note that the electronic structure and electron-phonon mass-enhancement parameter for ZrTe$_2$ are nearly identical to TiSe$_2$ \cite{zunger1978, fang1997, bianco2015, das2015}. TiSe$_2$ is composed by Se-4p hole-pockets at the center of its first Brillouin zone ($\Gamma$) and Ti-3$d$ electron-pockets located at L high symmetry-point with electron-phonon coupling constant $\approx 0.65$ \cite{das2015}, very similar to ZrTe$_2$. This compound is widely investigated, and it can be regarded as a prototype for the study of the connection between CDW and superconductivity in low-dimensional materials \cite{morosan2006, kusmartseva2009, morosan2010, rohwer2011, mohr2011, li2016, kitou2019}. The onset of a CDW phase transition in TiSe$_2$ is marked by an anomaly in the electrical resistivity at $T = 202$\,K that is quite similar to the 287\,K anomaly in Ni$_{0.04}$ZrTe$_2$ \cite{di1976}. It is noteworthy to point out that the physical nature of CDWs in general is controversial. Theoretical and experimental studies suggest that rather than CDWs, the phenomena being observed in the bulk properties can be the result of exciton condensation (electron-hole coupling) leading to insulating behavior \cite{pillo2000, cercellier2007}, Jahn-Teller effect (electron-phonon coupling) \cite{rossnagel2002} or hybrid exciton-phonon modes \cite{van2010, watanabe2015}.

In the light of the similarity of the electronic ground states and electron-phonon coupling strength for ZrTe$_2$ and TiSe$_2$, we argue that possibly ZrTe$_2$ is able to harbor a hidden electronic CDW phase transition through an exciton-phonon-driven mechanism. Tuning the density of states at the Fermi level, the Couloumb electron-hole interaction can give rise to electron-hole bound states, i.e. exciton quasiparticles. If the binding energy of the electron-hole pair is greater than the energy difference between the maximum of the valence band and the minimum of the conduction band, the system becomes energetically unstable, undergoing an excitonic condensate with phase coherence and a periodicity defined by the wave-vector that connects the valence and conduction sheets \cite{cercellier2007}. The exciton formation will be favored in systems with low density of states and compensated electron- and hole-type carriers at the Fermi surface, as shown in the case of Ni$_{0.04}$ZrTe$_2$. 

\begin{figure*}[t]
    \centering
	\includegraphics[width=.7\linewidth]{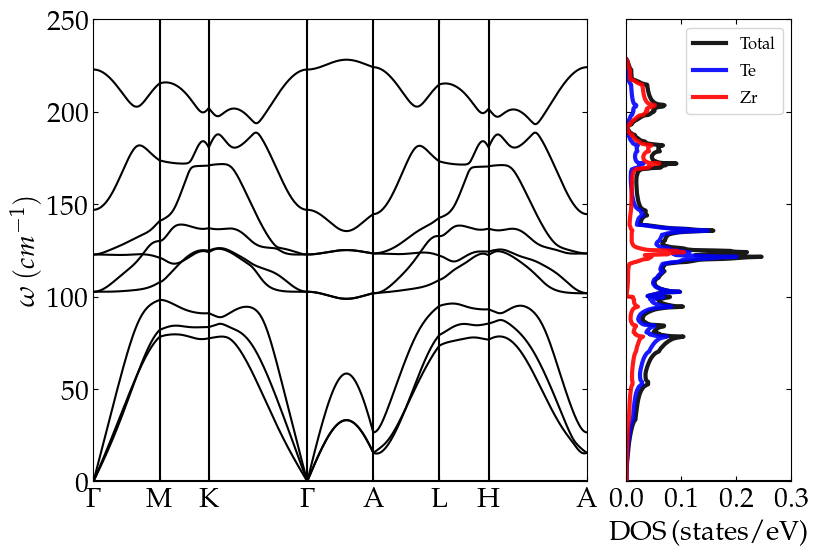}
	\caption{ZrTe$_2$ phonon dispersion and phonon DOS. There is a narrow softening of the acoustic phonon branches along $\Gamma$--A. Ni intercalation could enhance such Kohn anomaly due to increased screening effects on the Fermi surface, leading to a structural phase transition and to a lower symmetry phase.}
	\label{fig:phonon}
\end{figure*}

However, the manifestation of electron-phonon superconductivity suggests that the parent CDW state could be mediated by a combination of phonons and excitonic fluctuations. The phonon dispersion of ZrTe$_2$ with van der Waals non-local corrections obtained from Density Functional Perturbation Theory (DFPT) \cite{Baroni1987,Gonze1995} is shown in Fig. \ref{fig:phonon}. A softening of the acoustic phonon branches occurs along $\Gamma$--A. It is well-known that Kohn anomalies \cite{kohn1959} could induce a Peierls distortion, where a structural phase transition takes place due to the softening down to zero of a particular phonon mode \cite{duong2015}. Within the hybridized exciton-phonon scenario, increasing the exciton binding energy would significantly enhance the softening of the phonon spectrum, which promote structural instabilities \cite{van2010}.

Therefore, it is plausible that ZrTe$_2$ could harbor hidden CDW instabilities which can be brought forward by small Ni-doping levels, as suggested by the observed anomalies in $\rho(T)$ and C$_p(T)$ near 287\,K. The band-structure calculation suggests that the intercalation of Ni in the ZrTe$_2$ structure fine-tunes the Fermi surface topography by enhancing the Te-$p$ contribution at the Fermi level, creating a delicate balance between the electron-phonon and electron-hole interactions necessary for the subtle competition between superconductivity and CDW order.

\subsection{Band structure topology}

Additionally, due to the strong SOC of the Te-p wavefunction, the electronic dispersion of ZrTe$_2$ is heavily affected. It is possible to observe electronic broken degeneracies over the entire extent of first Brillouin zone in the vicinity of the Fermi level with SOC [see Fig. \ref{fig:band}(a)-(b)]. However, the double group irreductible representations of the band-crossings along $\Gamma$--$A$ direction (out-of-plane), two-fold degenerate due to inversion- and time-reversal-symmetries, are protected by $C_3$ rotational-symmetry of the $D_{3d}$ point-group along the $k_z$ axis \cite{ferreira2021}, including spin degree-of-freedom. Thus, the linear band crossings above $E_F$ along $\Gamma$--A are allowed, which result in Dirac cones that can be observed in the electronic band-structure.  
\begin{figure*}[t]
	\centering
	\subfloat[][]{\includegraphics[width=.3\columnwidth]{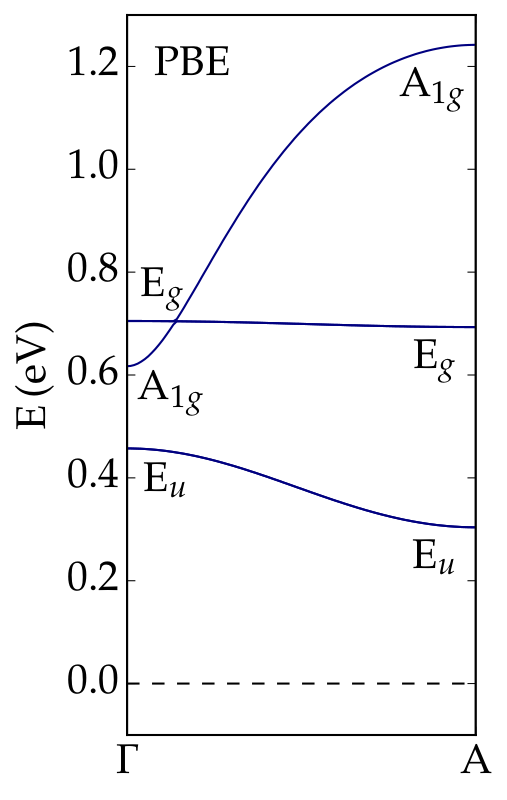}}
	\subfloat[][]{\includegraphics[width=.3\columnwidth]{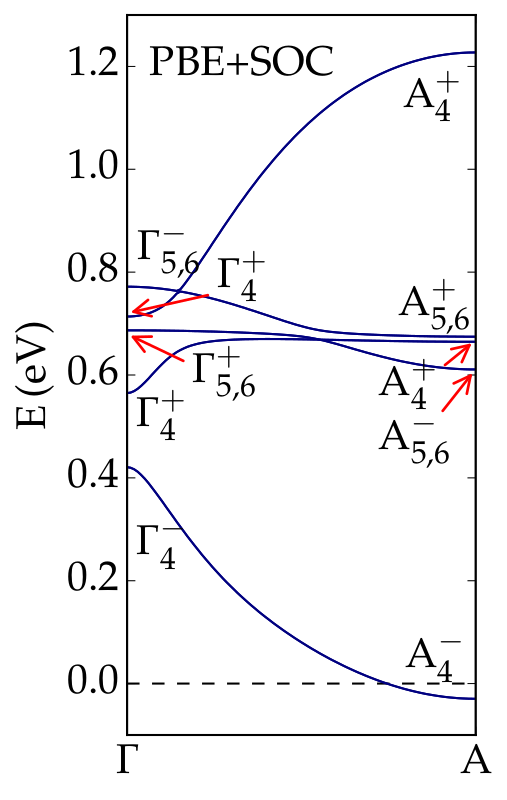}}
	\caption{Irreducible representations and parity analysis along $\Gamma$--A direction of ZrTe$_2$ (a) without and (b) with SOC effects.}
	\label{fig:IR}
\end{figure*}

The group analysis of the electronic dispersion along the $\Gamma$--A high-symmetry line, without (a) and with (b) SOC effects, is shown in Fig. \ref{fig:IR}. In the absence of spin-orbit coupling, the bands with E$_u$ and E$_g$ representations are fourfold degenerate and the E$_g$ states have a low dispersion along the out-of-plane direction. Equally important, there is a band-crossing between the E$_g$ and A$_{1g}$ bands, occurring at approximately 0.7\,eV above the Fermi level. Including the SOC effects, the degeneracy is broken, abruptly changing the electronic dispersion. The E$_u$ will give rise to the double group R$^{\pm}_{4}$ representation, while the E$_g$ derived states will result in low dispersive electronic-states with R$^{\pm}_{5,6}$ representation. As a result, we will have the formation of two linear band-crossings that resemble tilted Dirac cones between 0.6\,eV and 0.8\,eV derived from the crossings between R$_{5,6}$ and R$_{4}$ representations. At the same time, SOC-broken degeneracy at the E$_g$ states results in parity inversion, promoting a topologically non-trivial state. 

Therefore, ZrTe$_2$ is much more likely to reach a non-trivial topological type-II Dirac semimetal state, in contrast to recent studies claiming a topological nodal line state \cite{tian2020}, so that a comprehensive experimental characterization of the topological properties is still in order, and the role of Ni doping needs to be explored further. Due to the Dirac cones' energy it is very unlikely that the emergent Dirac-type quasiparticles will contribute somehow to the superconductivity or CDW order in ZrTe$_2$. However, strain or intercalation engineering could drive the system into a region in the parameter space where the Dirac cones are in the close vicinity of the Fermi level \cite{ferreira2021}. This is an issue where we hope our work could motivate further experiments, as well as the development of new models.

\section{Conclusions}

\label{sec:conclusions}

In summary, we have investigated the normal and superconducting phases of Ni$_{0.04}$ZrTe$_2$ single crystals by electrical resistivity, specific heat, and magnetization measurements. The electrical resistivity and magnetization data show the onset of superconductivity with T$_c \approx 4.1$\,K. However, the specific heat data near T$_c$ do not show the type of anomaly seen in second order phase transitions in bulk materials, thus raising a question still open about the nature of superconductivity in Ni$_{0.04}$ZrTe$_2$. Resistivity and specific heat data show clear features centered near 287\,K that are suggestive of a CDW. Therefore, Ni$_{0.04}$ZrTe$_2$ adds to the small list of materials where superconductivity and CDW compete for the Fermi surface. The temperature dependence of the upper critical field and the normalized superfluid density can be consistently fitted with a two-gap model, suggesting that there are two superconducting order parameters. 

The DFT calculations support the experimental findings remarkably well, and they suggest that ZrTe$_2$ could reach a non-trivial topological type-II Dirac semimetallic state via strain or intercalation engineering. The electronic structure calculations indicate that the main result for the inclusion of Ni in the van der Waals gap is the stabilization of the transferred charge to the Te sites, gradually increasing both the total density of states at the Fermi energy and the contribution of the Te-p manifold at the superconducting pairing, consequently driving the system to a compensated electronic regime where electron and holes contribute equally at the Fermi surface, thus favoring the emergence of the CDW order. Additionally, the electron-phonon coupling strength combined with the Fermi surface topography and the possible softening of the acoustic phonon modes in ZrTe$_2$ are consistent with an electron-phonon multiband superconductivity and an exciton-phonon induced CDW condensation.

Given the uniqueness of the electronic structure in Ni-doped ZrTe$_2$, and the coexistence of superconductivity and CDW ground states, such compounds hold the promise to be excellent probing grounds in the investigation of coherent and non-trivial quantum states.

\section*{Acknowledgments}

We gratefully acknowledge the financial support of S\~ao Paulo Research Foundation (FAPESP) under Grants 2018/10835-6, 2018/08819-2, 2019/14359-7, 2019/05005-7, and 2020/08258-0. This study was also financed by the Coordena\c{c}\~ao de Aperfei\c{c}oamento de Pessoal de N\'{\i}vel Superior (CAPES) - Brazil - Finance Code 001. 

This research was partially conducted using high-performance computing resources made available by the Superintend\^encia de Tecnologia da Informa\c c\~ao (STI), Universidade de S\~ao Paulo. The authors also acknowledge the National Laboratory for Scientific Computing (LNCC/MCTI, Brazil) for providing HPC resources of the SDumont supercomputer, which have contributed to the research results reported within this paper. 

The authors would like to thank Michel L. M. dos Santos for helpful discussions about phonon calculations.

\section*{Data Availability}

The data that support the findings of this study are available from the corresponding author upon request. Please, do not hesitate to get in touch in case you have questions or want more information.

%\section*{References}

\begin{singlespace}
%\bibliographystyle{scriptamat}
%\bibliography{refs}

\begin{thebibliography}{109}
	\providecommand{\natexlab}[1]{#1}
	
	\bibitem[{Manzeli et~al.(2017)Manzeli, Ovchinnikov, Pasquier, Yazyev, and
		Kis}]{manzeli2017}
	S.~Manzeli, D.~Ovchinnikov, D.~Pasquier, O.~V. Yazyev, A.~Kis.
	\newblock Nature Reviews Materials 2 (2017) 17033.
	
	\bibitem[{Wang et~al.(2012)Wang, Kalantar-Zadeh, Kis, Coleman, and
		Strano}]{wang2012}
	Q.~H. Wang, K.~Kalantar-Zadeh, A.~Kis, J.~N. Coleman, M.~S. Strano.
	\newblock Nature Nanotechnology 7 (2012) 699--712.
	
	\bibitem[{Chhowalla et~al.(2013)Chhowalla, Shin, Eda, Li, Loh, and
		Zhang}]{chhowalla2013}
	M.~Chhowalla, H.~S. Shin, G.~Eda, L.-J. Li, K.~P. Loh, H.~Zhang.
	\newblock Nature Chemistry 5 (2013) 263--275.
	
	\bibitem[{Neto(2001)}]{neto2001}
	A.~H.~C. Neto.
	\newblock Physical Review Letters 86 (2001) 4382.
	
	\bibitem[{Valla et~al.(2004)Valla, Fedorov, Johnson et~al.}]{valla2004}
	T.~Valla, A.~V. Fedorov, P.~D. Johnson, et~al.
	\newblock Physical Review Letters 92 (2004) 086401.
	
	\bibitem[{Morosan et~al.(2006)Morosan, Zandbergen, Dennis et~al.}]{morosan2006}
	E.~Morosan, H.~W. Zandbergen, B.~S. Dennis, et~al.
	\newblock Nature Physics 2 (2006) 544--550.
	
	\bibitem[{Sipos et~al.(2008)Sipos, Kusmartseva, Akrap, Berger, Forr{\'o}, and
		Tuti{\v{s}}}]{sipos2008}
	B.~Sipos, A.~F. Kusmartseva, A.~Akrap, H.~Berger, L.~Forr{\'o}, E.~Tuti{\v{s}}.
	\newblock Nature Materials 7 (2008) 960--965.
	
	\bibitem[{Zhou et~al.(2016)Zhou, Yuan, Jiang, and Law}]{zhou2016}
	B.~T. Zhou, N.~F. Yuan, H.-L. Jiang, K.~T. Law.
	\newblock Physical Review B 93 (2016) 180501.
	
	\bibitem[{Bhoi et~al.(2016)Bhoi, Khim, Nam et~al.}]{bhoi2016}
	D.~Bhoi, S.~Khim, W.~Nam, et~al.
	\newblock Scientific Reports 6 (2016) 1--10.
	
	\bibitem[{Han et~al.(2018)Han, Duong, Keum, Yun, and Lee}]{han2018}
	G.~H. Han, D.~L. Duong, D.~H. Keum, S.~J. Yun, Y.~H. Lee.
	\newblock Chemical Reviews 118 (2018) 6297--6336.
	
	\bibitem[{Zibouche et~al.(2014)Zibouche, Kuc, Musfeldt, and
		Heine}]{zibouche2014}
	N.~Zibouche, A.~Kuc, J.~Musfeldt, T.~Heine.
	\newblock Annalen der Physik 526 (2014) 395--401.
	
	\bibitem[{Xia et~al.(2014)Xia, Wang, Xiao, Dubey, and
		Ramasubramaniam}]{xia2014}
	F.~Xia, H.~Wang, D.~Xiao, M.~Dubey, A.~Ramasubramaniam.
	\newblock Nature Photonics 8 (2014) 899--907.
	
	\bibitem[{Yang et~al.(2015)Yang, Ji, and Jung}]{yang2015}
	E.~Yang, H.~Ji, Y.~Jung.
	\newblock The Journal of Physical Chemistry C 119 (2015) 26374--26380.
	
	\bibitem[{Wu et~al.(2017)Wu, Du, and Sun}]{wu2017}
	S.~Wu, Y.~Du, S.~Sun.
	\newblock Chemical Engineering Journal 307 (2017) 189--207.
	
	\bibitem[{Krasnok et~al.(2018)Krasnok, Lepeshov, and Al{\'u}}]{krasnok2018}
	A.~Krasnok, S.~Lepeshov, A.~Al{\'u}.
	\newblock Optics Express 26 (2018) 15972--15994.
	
	\bibitem[{David et~al.(2019)David, Rakyta, Korm{\'a}nyos, and
		Burkard}]{david2019}
	A.~David, P.~Rakyta, A.~Korm{\'a}nyos, G.~Burkard.
	\newblock Physical Review B 100 (2019) 085412.
	
	\bibitem[{Lucatto et~al.(2019)Lucatto, Koda, Bechstedt, Marques, and
		Teles}]{lucatto2019}
	B.~Lucatto, D.~S. Koda, F.~Bechstedt, M.~Marques, L.~K. Teles.
	\newblock Physical Review B 100 (2019) 121406.
	
	\bibitem[{Wagner et~al.(2008)Wagner, Morosan, Hor et~al.}]{wagner2008}
	K.~E. Wagner, E.~Morosan, Y.~S. Hor, et~al.
	\newblock Physical Review B 78 (2008) 104520.
	
	\bibitem[{Morosan et~al.(2010)Morosan, Wagner, Zhao et~al.}]{morosan2010}
	E.~Morosan, K.~E. Wagner, L.~L. Zhao, et~al.
	\newblock Physical Review B 81 (2010) 094524.
	
	\bibitem[{Kiswandhi et~al.(2013)Kiswandhi, Brooks, Cao et~al.}]{kiswandhi2013}
	A.~Kiswandhi, J.~S. Brooks, H.~B. Cao, et~al.
	\newblock Physical Review B 87 (2013) 121107.
	
	\bibitem[{Chang et~al.(2016)Chang, Chen, Bian et~al.}]{chang2016}
	T.-R. Chang, P.-J. Chen, G.~Bian, et~al.
	\newblock Physical Review B 93 (2016) 245130.
	
	\bibitem[{Guzman et~al.(2017)Guzman, Onofrio, and Strachan}]{guzman2017}
	D.~M. Guzman, N.~Onofrio, A.~Strachan.
	\newblock Journal of Applied Physics 121 (2017) 055703.
	
	\bibitem[{Kitou et~al.(2019)Kitou, Nakano, Kobayashi et~al.}]{kitou2019}
	S.~Kitou, A.~Nakano, S.~Kobayashi, et~al.
	\newblock Physical Review B 99 (2019) 104109.
	
	\bibitem[{De~Boer and Cordfunke(1997)}]{de1997}
	R.~De~Boer, E.~H.~P. Cordfunke.
	\newblock Journal of Alloys and Compounds 259 (1997) 115--121.
	
	\bibitem[{Reshak and Auluck(2004)}]{reshak2004}
	A.~H. Reshak, S.~Auluck.
	\newblock Physica B: Condensed Matter 353 (2004) 230--237.
	
	\bibitem[{Kar et~al.(2020)Kar, Chatterjee, Harnagea et~al.}]{kar2020}
	I.~Kar, J.~Chatterjee, L.~Harnagea, et~al.
	\newblock Physical Review B 101 (2020) 165122.
	
	\bibitem[{Tsipas et~al.(2018)Tsipas, Tsoutsou, Fragkos et~al.}]{tsipas2018}
	P.~Tsipas, D.~Tsoutsou, S.~Fragkos, et~al.
	\newblock ACS nano 12 (2018) 1696--1703.
	
	\bibitem[{Tian et~al.(2020)Tian, Ghassemi, and Ross~Jr}]{tian2020}
	Y.~Tian, N.~Ghassemi, J.~H. Ross~Jr.
	\newblock Physical Review B 102 (2020) 165149.
	
	\bibitem[{Machado et~al.(2017)Machado, Baptista, De~Lima et~al.}]{machado2017}
	A.~J.~S. Machado, N.~P. Baptista, B.~S. De~Lima, et~al.
	\newblock Physical Review B 95 (2017) 144505.
	
	\bibitem[{Ferreira et~al.(2021)Ferreira, Manesco, Dorini et~al.}]{ferreira2021}
	P.~P. Ferreira, A.~L. Manesco, T.~T. Dorini, et~al.
	\newblock Physical Review B 103 (2021) 125134.
	
	\bibitem[{de~Lima et~al.(2018)de~Lima, de~Cassia, Santos et~al.}]{de2018}
	B.~S. de~Lima, R.~R. de~Cassia, F.~B. Santos, et~al.
	\newblock Solid State Communications 283 (2018) 27--31.
	
	\bibitem[{Kraus and Nolze(1996)}]{kraus1996}
	W.~Kraus, G.~Nolze.
	\newblock Journal of Applied Crystallography 29 (1996) 301--303.
	
	\bibitem[{Zhang et~al.(2020)Zhang, Muhammad, Wang et~al.}]{zhang2020}
	B.~Zhang, Z.~Muhammad, P.~Wang, et~al.
	\newblock The Journal of Physical Chemistry C 124 (2020) 16561--16567.
	
	\bibitem[{Bunting et~al.(1969)Bunting, Ashworth, and Steeple}]{bunting1969}
	J.~Bunting, T.~Ashworth, H.~Steeple.
	\newblock Cryogenics 9 (1969) 385--386.
	
	\bibitem[{qua(2014)}]{quantum}
	Application Note 1085-152.
	\newblock Quantum Design, rev.b0 ed. (2014).
	
	\bibitem[{Hohenberg and Kohn(1964)}]{hohenberg1964}
	P.~Hohenberg, W.~Kohn.
	\newblock Physical Review 136 (1964) B864.
	
	\bibitem[{Kohn and Sham(1965)}]{kohn1965}
	W.~Kohn, L.~J. Sham.
	\newblock Physical Review 140 (1965) A1133.
	
	\bibitem[{Dal~Corso(2014)}]{dal2014}
	A.~Dal~Corso.
	\newblock Computational Materials Science 95 (2014) 337--350.
	
	\bibitem[{Giannozzi et~al.(2009)Giannozzi, Baroni, Bonini
		et~al.}]{giannozzi2009}
	P.~Giannozzi, S.~Baroni, N.~Bonini, et~al.
	\newblock Journal of physics: Condensed matter 21 (2009) 395502.
	
	\bibitem[{Giannozzi et~al.(2017)Giannozzi, Andreussi, Brumme
		et~al.}]{giannozzi2017}
	P.~Giannozzi, O.~Andreussi, T.~Brumme, et~al.
	\newblock Journal of Physics: Condensed Matter 29 (2017) 465901.
	
	\bibitem[{Kokalj(1999)}]{kokalj1999}
	A.~Kokalj.
	\newblock Journal of Molecular Graphics and Modelling 17 (1999) 176--179.
	
	\bibitem[{Kawamura(2019)}]{kawamura2019}
	M.~Kawamura.
	\newblock Computer Physics Communications 239 (2019) 197--203.
	
	\bibitem[{Perdew et~al.(1996)Perdew, Burke, and Ernzerhof}]{perdew1996}
	J.~P. Perdew, K.~Burke, M.~Ernzerhof.
	\newblock Physical Review Letters 77 (1996) 3865.
	
	\bibitem[{Thonhauser et~al.(2007)Thonhauser, Cooper, Li, Puzder, Hyldgaard, and
		Langreth}]{thonhauser2007}
	T.~Thonhauser, V.~R. Cooper, S.~Li, A.~Puzder, P.~Hyldgaard, D.~C. Langreth.
	\newblock Physical Review B 76 (2007) 125112.
	
	\bibitem[{Langreth et~al.(2009)Langreth, Lundqvist, Chakarova-K{\"a}ck
		et~al.}]{langreth2009}
	D.~Langreth, B.~I. Lundqvist, S.~D. Chakarova-K{\"a}ck, et~al.
	\newblock Journal of Physics: Condensed Matter 21 (2009) 084203.
	
	\bibitem[{Sabatini et~al.(2012)Sabatini, K{\"u}{\c{c}}{\"u}kbenli, Kolb,
		Thonhauser, and De~Gironcoli}]{sabatini2012}
	R.~Sabatini, E.~K{\"u}{\c{c}}{\"u}kbenli, B.~Kolb, T.~Thonhauser,
	S.~De~Gironcoli.
	\newblock Journal of Physics: Condensed Matter 24 (2012) 424209.
	
	\bibitem[{Hamada(2014)}]{hamada2014}
	I.~Hamada.
	\newblock Physical Review B 89 (2014) 121103.
	
	\bibitem[{Thonhauser et~al.(2015)Thonhauser, Zuluaga, Arter, Berland,
		Schr{\"o}der, and Hyldgaard}]{thonhauser2015}
	T.~Thonhauser, S.~Zuluaga, C.~Arter, K.~Berland, E.~Schr{\"o}der, P.~Hyldgaard.
	\newblock Physical review letters 115 (2015) 136402.
	
	\bibitem[{Berland et~al.(2015)Berland, Cooper, Lee et~al.}]{berland2015}
	K.~Berland, V.~R. Cooper, K.~Lee, et~al.
	\newblock Reports on Progress in Physics 78 (2015) 066501.
	
	\bibitem[{Marzari et~al.(1999)Marzari, Vanderbilt, De~Vita, and
		Payne}]{marzari1999}
	N.~Marzari, D.~Vanderbilt, A.~De~Vita, M.~Payne.
	\newblock Physical review letters 82 (1999) 3296.
	
	\bibitem[{Okhotnikov et~al.(2016)Okhotnikov, Charpentier, and
		Cadars}]{okhotnikov2016}
	K.~Okhotnikov, T.~Charpentier, S.~Cadars.
	\newblock Journal of cheminformatics 8 (2016) 1--15.
	
	\bibitem[{Berthier et~al.(1976)Berthier, Molini{\'e}, and
		J{\'e}rome}]{berthier1976}
	C.~Berthier, P.~Molini{\'e}, D.~J{\'e}rome.
	\newblock Solid State Communications 18 (1976) 1393--1395.
	
	\bibitem[{Craven and Meyer(1977)}]{craven1977}
	R.~A. Craven, S.~F. Meyer.
	\newblock Physical Review B 16 (1977) 4583.
	
	\bibitem[{Yang et~al.(2014)Yang, Wang, Liu et~al.}]{yang2014}
	J.~Yang, W.~Wang, Y.~Liu, et~al.
	\newblock Applied Physics Letters 105 (2014) 063109.
	
	\bibitem[{Zhu et~al.(2015)Zhu, Cao, Zhang, Plummer, and Guo}]{zhu2015}
	X.~Zhu, Y.~Cao, J.~Zhang, E.~Plummer, J.~Guo.
	\newblock Proceedings of the National Academy of Sciences 112 (2015)
	2367--2371.
	
	\bibitem[{Kolincio et~al.(2017)Kolincio, Roman, Winiarski, Strychalska-Nowak,
		and Klimczuk}]{kolincio2017}
	K.~K. Kolincio, M.~Roman, M.~J. Winiarski, J.~Strychalska-Nowak, T.~Klimczuk.
	\newblock Physical Review B 95 (2017) 235156.
	
	\bibitem[{Pfuner et~al.(2010)Pfuner, Lerch, Chu, Kuo, Fisher, and
		Degiorgi}]{pfuner2010}
	F.~Pfuner, P.~Lerch, J.-H. Chu, H.-H. Kuo, I.~Fisher, L.~Degiorgi.
	\newblock Physical Review B 81 (2010) 195110.
	
	\bibitem[{Denholme et~al.(2017)Denholme, Yukawa, Tsumura et~al.}]{denholme2017}
	S.~J. Denholme, A.~Yukawa, K.~Tsumura, et~al.
	\newblock Scientific Reports 7 (2017) 45217.
	
	\bibitem[{Ren et~al.(2021)Ren, Han, Fan et~al.}]{ren2021}
	M.-Q. Ren, S.~Han, J.-Q. Fan, et~al.
	\newblock arXiv preprint arXiv:2102.07915  (2021).
	
	\bibitem[{Zunger and Freeman(1978)}]{zunger1978}
	A.~Zunger, A.~J. Freeman.
	\newblock Physical Review B 17 (1978) 1839.
	
	\bibitem[{Fang et~al.(1997)Fang, De~Groot, and Haas}]{fang1997}
	C.~M. Fang, R.~A. De~Groot, C.~Haas.
	\newblock Physical Review B 56 (1997) 4455.
	
	\bibitem[{Bianco et~al.(2015)Bianco, Calandra, and Mauri}]{bianco2015}
	R.~Bianco, M.~Calandra, F.~Mauri.
	\newblock Physical Review B 92 (2015) 094107.
	
	\bibitem[{Das and Dolui(2015)}]{das2015}
	T.~Das, K.~Dolui.
	\newblock Physical Review B 91 (2015) 094510.
	
	\bibitem[{Werthamer et~al.(1966)Werthamer, Helfand, and Hohenberg}]{WHH}
	N.~R. Werthamer, E.~Helfand, P.~C. Hohenberg.
	\newblock Phys. Rev. 147 (1966) 295--302.
	
	\bibitem[{Lyard et~al.(2002)Lyard, Samuely, Szabo et~al.}]{lyard2002}
	L.~Lyard, P.~Samuely, P.~Szabo, et~al.
	\newblock Physical Review B 66 (2002) 180502.
	
	\bibitem[{Hunte et~al.(2008)Hunte, Jaroszynski, Gurevich et~al.}]{hunte2008}
	F.~Hunte, J.~Jaroszynski, A.~Gurevich, et~al.
	\newblock Nature 453 (2008) 903--905.
	
	\bibitem[{Lei et~al.(2012)Lei, Graf, Hu et~al.}]{lei2012}
	H.~Lei, D.~Graf, R.~Hu, et~al.
	\newblock Physical Review B 85 (2012) 094515.
	
	\bibitem[{Li et~al.(2018)Li, Gu, Chen et~al.}]{li2018}
	Y.~Li, Q.~Gu, C.~Chen, et~al.
	\newblock Proceedings of the National Academy of Sciences 115 (2018)
	9503--9508.
	
	\bibitem[{Santos et~al.(2018)Santos, Correa, de~Lima et~al.}]{santos2018}
	F.~B. Santos, L.~E. Correa, B.~S. de~Lima, et~al.
	\newblock Physics Letters A 382 (2018) 1065--1068.
	
	\bibitem[{Shang et~al.(2019)Shang, Amon, Kasinathan et~al.}]{shang2019}
	T.~Shang, A.~Amon, D.~Kasinathan, et~al.
	\newblock New Journal of Physics 21 (2019) 073034.
	
	\bibitem[{Xu et~al.(2019)Xu, Li, Feng et~al.}]{xu2019}
	C.~Xu, B.~Li, J.~Feng, et~al.
	\newblock Physical Review B 100 (2019) 134503.
	
	\bibitem[{Majumdar et~al.(2020)Majumdar, VanGennep, Brisbois
		et~al.}]{majumdar2020}
	A.~Majumdar, D.~VanGennep, J.~Brisbois, et~al.
	\newblock Physical Review Materials 4 (2020) 084005.
	
	\bibitem[{Gurevich(2003)}]{gurevich2003}
	A.~Gurevich.
	\newblock Physical Review B 67 (2003) 184515.
	
	\bibitem[{Silaev and Babaev(2011)}]{silaev2011}
	M.~Silaev, E.~Babaev.
	\newblock Phys. Rev. B 84 (2011) 094515.
	
	\bibitem[{Silaev and Babaev(2012)}]{silaev2012}
	M.~Silaev, E.~Babaev.
	\newblock Phys. Rev. B 85 (2012) 134514.
	
	\bibitem[{Cavalcanti et~al.(2020)Cavalcanti, Saraiva, Aguiar, Vagov, Croitoru,
		and Shanenko}]{cavalcanti2020}
	P.~J.~F. Cavalcanti, T.~T. Saraiva, J.~A. Aguiar, A.~Vagov, M.~Croitoru, A.~A.
	Shanenko.
	\newblock Journal of Physics: Condensed Matter 32 (2020) 455702.
	
	\bibitem[{Angst et~al.(2002)Angst, Puzniak, Wisniewski et~al.}]{angst2002}
	M.~Angst, R.~Puzniak, A.~Wisniewski, et~al.
	\newblock Physical Review Letters 88 (2002) 167004.
	
	\bibitem[{Ren et~al.(2008)Ren, Wang, Luo, Yang, Shan, and Wen}]{ren2008}
	C.~Ren, Z.-S. Wang, H.-Q. Luo, H.~Yang, L.~Shan, H.-H. Wen.
	\newblock Physical Review Letters 101 (2008) 257006.
	
	\bibitem[{Kim et~al.(2002)Kim, Skinta, Lemberger et~al.}]{kim2002}
	M.-S. Kim, J.~A. Skinta, T.~R. Lemberger, et~al.
	\newblock Physical Review B 66 (2002) 064511.
	
	\bibitem[{Chandrasekhar and Einzel(1993)}]{chandrasekhar1993}
	B.~S. Chandrasekhar, D.~Einzel.
	\newblock Annalen der Physik 505 (1993) 535--546.
	
	\bibitem[{Carrington and Manzano(2003)}]{carrington2003}
	A.~Carrington, F.~Manzano.
	\newblock Physica C: Superconductivity 385 (2003) 205--214.
	
	\bibitem[{Prozorov and Giannetta(2006)}]{prozorov2006}
	R.~Prozorov, R.~W. Giannetta.
	\newblock Superconductor Science and Technology 19 (2006) R41.
	
	\bibitem[{Jishi and Alyahyaei(2008)}]{jishi2008}
	R.~A. Jishi, H.~M. Alyahyaei.
	\newblock Physical Review B 78 (2008) 144516.
	
	\bibitem[{McMillan(1968)}]{mcmillan1968}
	W.~L. McMillan.
	\newblock Physical Review 167 (1968) 331.
	
	\bibitem[{McMillan and Rowell(1965)}]{mcmillan1965}
	W.~L. McMillan, J.~M. Rowell.
	\newblock Physical Review Letters 14 (1965) 108.
	
	\bibitem[{Ferreira et~al.(2018)Ferreira, Santos, Machado, Petrilli, and
		Eleno}]{ferreira2018}
	P.~P. Ferreira, F.~B. Santos, A.~J.~S. Machado, H.~M. Petrilli, L.~T.~F. Eleno.
	\newblock Physical Review B 98 (2018) 045126.
	
	\bibitem[{Bennemann and Garland(1972)}]{bennemann1972}
	K.~H. Bennemann, J.~W. Garland.
	\newblock In AIP Conference Proceedings. American Institute of Physics, vol.~4,
	pp. 103--137.
	
	\bibitem[{Aoki et~al.(1996)Aoki, Sambongi, Levy, and Berger}]{aoki1996}
	Y.~Aoki, T.~Sambongi, F.~Levy, H.~Berger.
	\newblock Journal of the Physical Society of Japan 65 (1996) 2590--2593.
	
	\bibitem[{Floris et~al.(2007)Floris, Sanna, Massidda, and Gross}]{floris2007}
	A.~Floris, A.~Sanna, S.~Massidda, E.~K.~U. Gross.
	\newblock Physical Review B 75 (2007) 054508.
	
	\bibitem[{Singh(2013)}]{singh2013}
	D.~J. Singh.
	\newblock Physical Review B 88 (2013) 174508.
	
	\bibitem[{Singh(2015)}]{singh2015}
	D.~J. Singh.
	\newblock PloS One 10 (2015) e0123667.
	
	\bibitem[{Bhattacharyya et~al.(2020)Bhattacharyya, Ferreira, Santos
		et~al.}]{bhattacharyya2020}
	A.~Bhattacharyya, P.~P. Ferreira, F.~B. Santos, et~al.
	\newblock Physical Review Research 2 (2020) 022001.
	
	\bibitem[{Zhao et~al.(2020)Zhao, Lian, Zeng, Dai, Meng, and Ni}]{zhao2020}
	Y.~Zhao, C.~Lian, S.~Zeng, Z.~Dai, S.~Meng, J.~Ni.
	\newblock Physical Review B 101 (2020) 104507.
	
	\bibitem[{de~Faria et~al.(2021)de~Faria, Ferreira, Correa, Eleno, Torikachvili,
		and Machado}]{de2021}
	L.~R. de~Faria, P.~P. Ferreira, L.~E. Correa, L.~T.~F. Eleno, M.~S.
	Torikachvili, A.~J.~S. Machado.
	\newblock Superconductor Science and Technology  (2021).
	
	\bibitem[{Chen et~al.(2020)Chen, Zhu, and Shanenko}]{chen2020}
	Y.~Chen, H.~Zhu, A.~A. Shanenko.
	\newblock Physical Review B 101 (2020) 214510.
	
	\bibitem[{Kusmartseva et~al.(2009)Kusmartseva, Sipos, Berger, Forro, and
		Tuti{\v{s}}}]{kusmartseva2009}
	A.~F. Kusmartseva, B.~Sipos, H.~Berger, L.~Forro, E.~Tuti{\v{s}}.
	\newblock Physical Review Letters 103 (2009) 236401.
	
	\bibitem[{Rohwer et~al.(2011)Rohwer, Hellmann, Wiesenmayer et~al.}]{rohwer2011}
	T.~Rohwer, S.~Hellmann, M.~Wiesenmayer, et~al.
	\newblock Nature 471 (2011) 490--493.
	
	\bibitem[{M{\"o}hr-Vorobeva et~al.(2011)M{\"o}hr-Vorobeva, Johnson, Beaud
		et~al.}]{mohr2011}
	E.~M{\"o}hr-Vorobeva, S.~L. Johnson, P.~Beaud, et~al.
	\newblock Physical Review Letters 107 (2011) 036403.
	
	\bibitem[{Li et~al.(2016)Li, O’Farrell, Loh, Eda, {\"O}zyilmaz, and
		Neto}]{li2016}
	L.~Li, E.~O’Farrell, K.~Loh, G.~Eda, B.~{\"O}zyilmaz, A.~C. Neto.
	\newblock Nature 529 (2016) 185--189.
	
	\bibitem[{Di~Salvo et~al.(1976)Di~Salvo, Moncton, and Waszczak}]{di1976}
	F.~J. Di~Salvo, D.~E. Moncton, J.~V. Waszczak.
	\newblock Physical Review B 14 (1976) 4321.
	
	\bibitem[{Pillo et~al.(2000)Pillo, Hayoz, Berger, L{\'e}vy, Schlapbach, and
		Aebi}]{pillo2000}
	T.~Pillo, J.~Hayoz, H.~Berger, F.~L{\'e}vy, L.~Schlapbach, P.~Aebi.
	\newblock Physical Review B 61 (2000) 16213.
	
	\bibitem[{Cercellier et~al.(2007)Cercellier, Monney, Clerc
		et~al.}]{cercellier2007}
	H.~Cercellier, C.~Monney, F.~Clerc, et~al.
	\newblock Physical Review Letters 99 (2007) 146403.
	
	\bibitem[{Rossnagel et~al.(2002)Rossnagel, Kipp, and Skibowski}]{rossnagel2002}
	K.~Rossnagel, L.~Kipp, M.~Skibowski.
	\newblock Physical Review B 65 (2002) 235101.
	
	\bibitem[{van Wezel et~al.(2010)van Wezel, Nahai-Williamson, and
		Saxena}]{van2010}
	J.~van Wezel, P.~Nahai-Williamson, S.~S. Saxena.
	\newblock Physical Review B 81 (2010) 165109.
	
	\bibitem[{Watanabe et~al.(2015)Watanabe, Seki, and Yunoki}]{watanabe2015}
	H.~Watanabe, K.~Seki, S.~Yunoki.
	\newblock Physical Review B 91 (2015) 205135.
	
	\bibitem[{Baroni et~al.(1987)Baroni, Giannozzi, and Testa}]{Baroni1987}
	S.~Baroni, P.~Giannozzi, A.~Testa.
	\newblock Phys. Rev. Lett. 58 (1987) 1861--1864.
	
	\bibitem[{Gonze(1995)}]{Gonze1995}
	X.~Gonze.
	\newblock Phys. Rev. A 52 (1995) 1086--1095.
	
	\bibitem[{Kohn(1959)}]{kohn1959}
	W.~Kohn.
	\newblock Physical Review Letters 2 (1959) 393.
	
	\bibitem[{Duong et~al.(2015)Duong, Burghard, and Sch{\"o}n}]{duong2015}
	D.~L. Duong, M.~Burghard, J.~C. Sch{\"o}n.
	\newblock Physical Review B 92 (2015) 245131.
	
\end{thebibliography}
%\input{refs.tex}

\end{singlespace}

\end{document}